\documentclass[fleqn,usenatbib]{mnras}

\usepackage{newtxtext,newtxmath}
\usepackage{longtable}

\usepackage[T1]{fontenc}

\DeclareRobustCommand{\VAN}[3]{#2}
\let\VANthebibliography\thebibliography
\def\thebibliography{\DeclareRobustCommand{\VAN}[3]{##3}\VANthebibliography}

\usepackage{pdflscape}
\usepackage{graphicx}	
\usepackage{amsmath}	
\usepackage{natbib}
\usepackage{rotating}
\usepackage{multirow}
\usepackage{hyperref}
\usepackage{booktabs} 
\usepackage{float}
\usepackage{academicons}
\usepackage{xcolor}

\title[Blue Straggler Stars in Berkeley 18]{Blue Straggler Stars in Berkeley 18: A Multiwavelength Study of Their Physical Properties and Dynamical Evolution}

\author[D. Bisht et al.]{
D. Bisht\orcid{0000-0002-8988-8434}$^{1}$\thanks{E-mail: devendrabisht297@gmail.com},
Ing-Guey Jiang\orcid{0000-0001-7359-3300}$^{2}$\thanks{E-mail:jiang@phys.nthu.edu.tw},
K. Belwal$^{1}$\thanks{E-mail:kuldeepbelwal1997@gmail.com },
D. C. \c{C}ınar\orcid{0000-0001-7940-3731}$^{3}$\thanks{E-mail:denizcennetcinar@gmail.com},
Alok Durgapal$^{4}$,
Shraddha Biswas\orcid{0009-0003-8446-4557}$^{1}$,\newauthor
A. Raj$^{1}$,
Geeta Rangwal\orcid{0000-0002-6373-770X}$^{5}$,
M. S. Bisht$^{1}$,
and M. Manu$^{6}$
\\
$^{1}$Indian Centre For Space Physics, 466 Barakhola, Singabari Road, Netai Nagar, Kolkata, West Bengal 700099, India\\
$^{2}$Department of Physics and Institute of Astronomy, National Tsing-Hua University, Hsinchu 30013, Taiwan\\
$^{3}$Programme of Astronomy and Space Sciences, Institute of Graduate Studies in Science, Istanbul University, 34116 Istanbul, Turkey\\
$^{4}$Center of Advanced Study, Department of Physics, D. S. B. Campus, Kumaun University, Nainital 263002, India\\
$^{5}$Aryabhatta Research Institute of Observational Sciences, Manora Peak, Nainital 263129, India\\
$^{6}$School of Allied Sciences, Graphic Era Hill University, Bhimtal Campus, 263136, India
}

\date{Accepted XXX. Received YYY; in original form ZZZ}
\pubyear{\the\year}
\usepackage{lineno}
\begin{document}
\label{firstpage}
\pagerange{\pageref{firstpage}--\pageref{lastpage}}
\maketitle
\begin{abstract}

Berkeley~18 is an old open cluster in the outer Galactic disk that hosts a population of blue straggler stars (BSSs). We present a comprehensive multiwavelength analysis of its BSS population using \textit{Gaia} DR3 astrometry, optical--infrared photometry, and time-domain TESS observations. Using a Gaussian Mixture Model (GMM) in astrometric space, we identify 798 high-probability cluster members ($p > 0.7$). Isochrone fitting yields an age of $3.2 \pm 0.2$ Gyr and a heliocentric distance of $5.01^{+0.75}_{-0.55}$ kpc. We identify 24 BSS candidates above the main-sequence turn-off. Spectral energy distribution (SED) modelling reveals effective temperatures of $6000$--$8500$ K, radii of $1.4$--$5.7\,R_\odot$, and luminosities of $3.3$--$38\,L_\odot$, indicating a heterogeneous population spanning multiple evolutionary stages. The BSS population exhibits only a mild central concentration, with a low $A^{+}$ parameter and an extremely low stellar collision-rate proxy, implying weak mass segregation and an inefficient collisional channel. We find no significant photometric variability among the BSS candidates within TESS's sensitivity limits. Although WISE W3/W4 data initially suggested possible mid-infrared excesses, detailed image inspection and SPHEREx spectrophotometry indicate that these are caused by background contamination and blending, with no clear evidence of circumstellar dust. The structural parameters derived from King-profile fitting ($r_c = 6.91^{+0.91}_{-0.73}$ arcmin, $r_t = 13.23^{+0.44}_{-0.43}$ arcmin) indicate a dynamically evolved, low-density system. Together, these results suggest that dynamical interactions are inefficient in Berkeley~18 and that binary evolution is likely the dominant formation channel of BSSs.

\end{abstract}

\begin{keywords}
open clusters and associations: individual: Berkeley 18 – blue stragglers – binaries: general – Hertzsprung–Russell diagrams – stars: kinematics and dynamics
\end{keywords}

\section{Introduction} 

Star clusters are ideal laboratories for studying stellar and Galactic evolution, as their member stars share similar ages, distances, and chemical compositions. This makes them excellent testbeds for theories of star formation, stellar dynamics, and the chemical and dynamical evolution of the Milky Way. \citep[e.g.,][]{Friel2002, Jacobson2016, Spina2017, Casamiquela2019, Zhong2020, Chen2020}. One especially interesting group in clusters is blue straggler stars (BSSs). In the Hertzsprung--Russell diagram, BSSs are brighter and bluer than the main-sequence turn-off (MSTO), which suggests they are more massive than other stars of the same cluster age \citep{Sandage1953}. They are usually explained as main-sequence stars that gained mass through binary evolution (e.g., mass transfer or mergers) and/or close stellar encounters \citep{Qin2026, Li2023, Rain2024, Rain2021, Jadhav2021a, Ahumada2007}.

BSSs are generally explained by two main formation pathways. In the first path, mass transfer or merging in a binary system increases a star's mass, making it appear younger \citep{McCrea1964}. In the second path, close encounters in dense regions lead to collisions or dynamical mergers, which can also produce BSSs \citep{Hills1976, Davies1994}. These two processes are fundamentally different, but both help us understand how clusters evolve, how binaries change over time, and how stars interact \citep{Bailyn1995, Glebbeek2010, Wyse2020}. It remains difficult to determine which path is more important, especially in old and intermediate-age open clusters (OCs), where stellar density is lower than in globular clusters, yet binary evolution may still dominate.

Old and intermediate-age OCs in the outer Galactic disk are good sites for studying this problem. Their ages provide sufficient time for binary evolution, and their locations help us understand the structure and evolution of the broader outer Milky Way, as well as its kinematics. In this context, the cluster Berkeley~18 (also known as Berk~18, Be~18, OCl~408, and MWSC~0516), located toward the Galactic anticenter, is a highly valuable target. Using CCD ($BVI$) photometry, \citet{Kaluzny1997} estimated the cluster age to be comparable to, or slightly younger than, that of M~67 (NGC~2682), placed it at a heliocentric distance of $\sim 5.8$~kpc, and found strong reddening with $E(B-V) > 0.46$. Subsequently, \citet{Carraro1999} re-analysed Berkeley~18 using $BV$ CCD photometry and derived a somewhat older age of $\sim 5$~Gyr, a distance of $\sim 5.5$~kpc, and a reddening of $E(B-V) = 0.52$, establishing it as one of the oldest  OCs in the outer Galactic disk. Its large angular diameter of $\sim 26$ arcmin (about $\sim 44$~pc) and its evolved stellar population suggest that it is dynamically evolved and has survived for several Gyr in the outer disk. Furthermore, Berkeley 18 contains a rich population of BSS candidates, making it an ideal system for studying BSS formation in the low-density conditions typical of old OCs.

To help break the degeneracy between the BSS formation channels, a multi-faceted observational approach is required. Recent studies have demonstrated that analyzing the spectral energy distribution (SED) from the ultraviolet to the infrared is critical for uncovering the hidden signatures of BSS formation. For instance, UV excesses can reveal hot white dwarf companions, the direct remnants of mass transfer \citep[e.g.,][]{Sindhu2020, Jadhav2021b, Sheikh2024}, while infrared excesses can indicate the presence of circumstellar dust or accretion disks resulting from recent stellar mergers or ongoing binary interactions \citep[e.g.,][]{Kervella2022, Dattatrey2023}. However, recent studies have shown that mid-infrared excess detections in crowded fields are highly susceptible to blending and background contamination, requiring careful validation. Furthermore, a cluster's dynamical evolution is deeply coupled to its Galactic orbit. By modeling the orbital kinematics, we can probe how the cluster's interaction with the Galactic tidal field and its radial migration history influence mass segregation and the close stellar encounters that fuel BSS formation \citep[e.g.,][]{Tarricq2021, Vaidya2020}.

In this work, we present a comprehensive, multiwavelength study of the BSS population in Berkeley 18 to better constrain their evolutionary pathways. By combining Gaia DR3 astrometry for precise membership determination with multiwavelength SED modelling and infrared diagnostics, we aim to search for signatures of binary interaction and to characterise the physical properties of the BSS population. In addition, we incorporate TESS time-domain observations to investigate photometric variability, which can provide complementary constraints on binarity and pulsational behaviour where detectable. Coupled with Galactic orbital modelling, this integrated approach provides useful constraints on the interplay between long-term binary evolution, stellar variability, and the dynamical environment of open clusters. Berkeley 18, therefore, provides an opportunity to investigate both stellar and potential circumstellar signatures of binary evolution in an old open cluster.

The paper is organised as follows. Section~\ref{sec:data} describes the data used in this work. Section~\ref{sec:membership} explains the GMM-based membership selection. Section~\ref{sec:rdp} presents the structural parameters derived from radial density profiles. Section~\ref{sec:CMD} discusses the colour--magnitude diagrams and the spatial distribution of BSSs. Section~\ref{sec:sed} presents the SED analysis and the search for binarity. Section~\ref{sec:bss_origin} examines infrared excess and the cluster's dynamical environment, while Section~\ref{sec:varability} discusses photometric variability. Finally, Section~\ref{sec:conclusion} summarises the main results and presents our conclusions.

\section{Data Used}
\label{sec:data}

\subsection{Gaia DR3 Data}

The optical data used in this study are primarily based on the third data release of the Gaia mission \citep{GaiaCollaboration2023}, which provides high-precision astrometry and broad-band photometry in the $G$, $G_{BP}$, and $G_{RP}$ bands. To construct the working sample, we extracted all sources within a radius of 20~arcmin from the cluster centre, resulting in a total of 17,298 stars. To ensure reliable astrometric solutions, we applied a quality cut of \texttt{RUWE} $\leq 1.4$ \citep{Lindegren2021}. This selection includes both cluster and field stars, providing a representative view of the surrounding region.

\begin{figure*}
    \centering
    \includegraphics[width=0.35\linewidth]{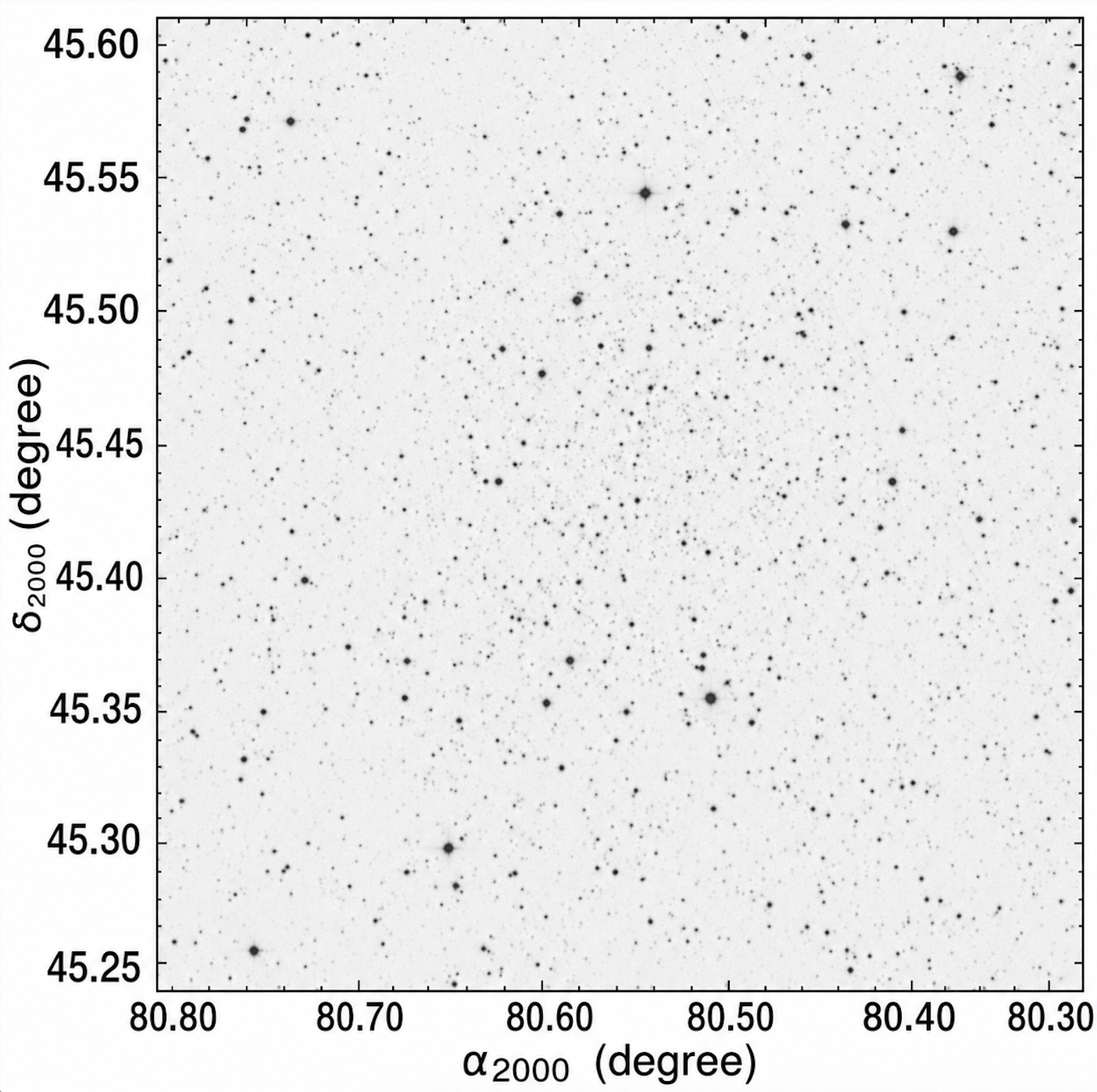}
    \includegraphics[width=0.35\linewidth]{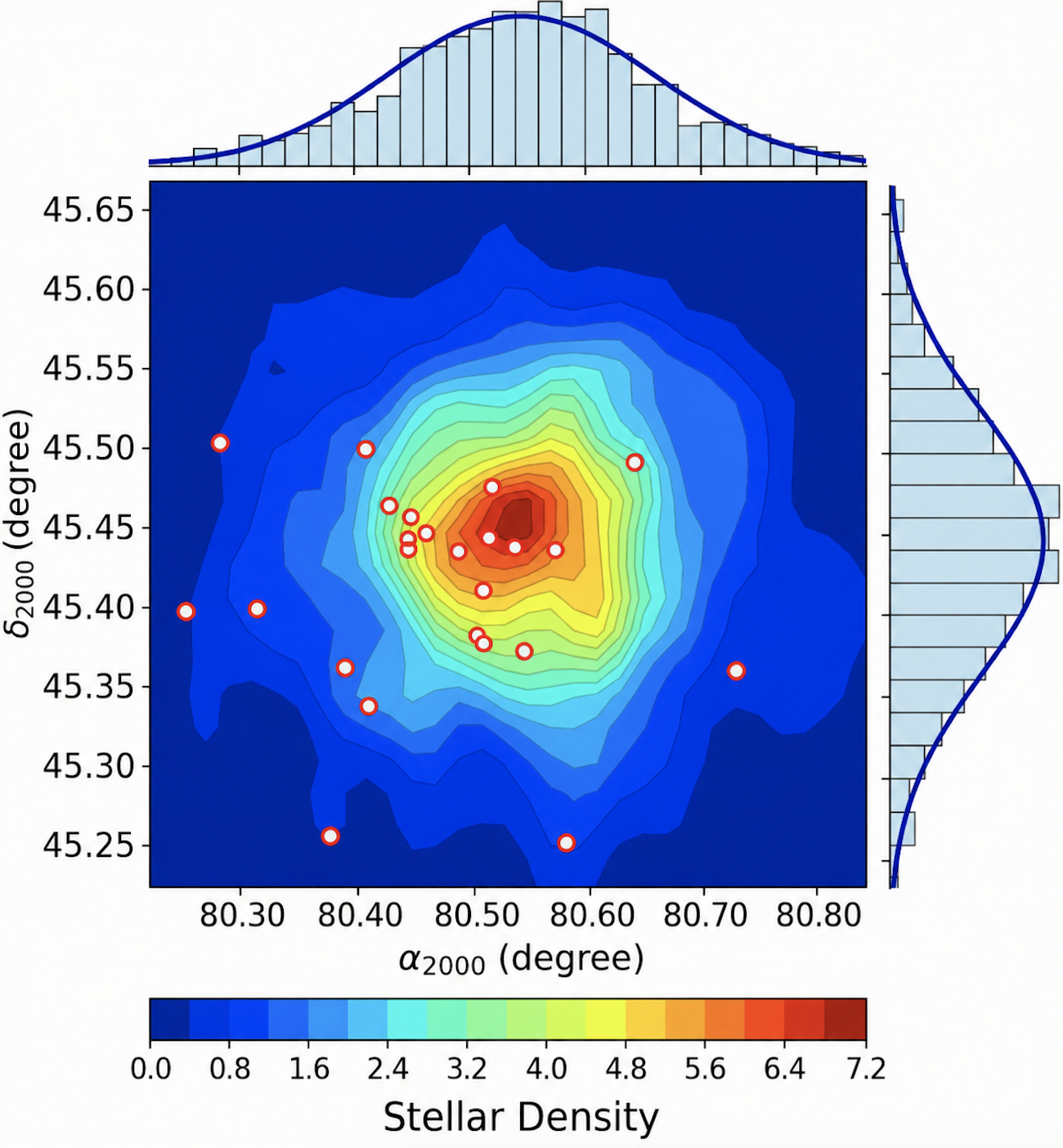}
    \caption{Left: Finder chart of the cluster based on DSS images. Right: Stellar surface density map derived from $Gaia$~DR3 data, with the colour scale representing the stellar density in units of stars~arcmin$^{-2}$. The overlaid contours highlight the cluster's central concentration and overall morphology.}
    \label{fig:charts}
\end{figure*}

The spatial distribution of the selected sources is shown in Figure~\ref{fig:charts}, which presents a finder chart (left panel) and a stellar surface density map (right panel). The density map reveals an overdensity associated with the cluster against the Galactic field. The finder chart is based on DSS images from the STScI Digitized Sky Survey\footnote{\url{https://archive.stsci.edu/cgi-bin/dss_form}}. This dataset forms the basis of the subsequent structural and photometric analysis. A magnitude limit is applied based on the photometric quality of the data, as discussed in Section~\ref{sec:phot_compl}.

\subsubsection{Photometric Completeness}\label{sec:phot_compl}

The photometric quality of the Gaia DR3 data is evaluated through the magnitude-dependent behaviour of the uncertainties (Table~\ref{tab:photometric_errors}). At bright magnitudes ($G \leq 17$ mag), the uncertainties remain nearly constant, with $\sigma_{\rm G} \sim 0.003$ mag and colour uncertainties below $\sim$0.01 mag. Toward fainter magnitudes, the uncertainties increase progressively. In the range $G = 18$-19 mag, the colour uncertainty rises to $\sim$0.04 mag, and becomes significantly larger beyond $G \approx 19$ mag, reaching $\sim$0.23 mag at $G = 20$-21 mag. For sources fainter than $G \approx 21$ mag, the uncertainties increase sharply, indicating a substantial degradation in photometric quality. Based on this behaviour, we adopt a conservative magnitude limit of $G = 20.1$ mag, below which the colour--magnitude diagram (CMD) remains well defined. Sources fainter than this limit are excluded from the analysis to avoid biases introduced by large photometric uncertainties.

\begin{table}
\centering
\small
\caption{Mean photometric uncertainties in Gaia DR3 $G$ magnitudes and $G_{\rm BP}-G_{\rm RP}$ colours for stars toward the cluster, reported as a function of $G$-band magnitude intervals.}
\label{tab:photometric_errors}
\begin{tabular}{c|ccc}
\hline
$G$ (mag) & $N$ & $\sigma_{{\rm G}}$ & $\sigma_{{G_{{\rm BP}}-G_{{\rm RP}}}}$ \\ \hline \hline
6--14 & 219 & 0.0028 & 0.0053 \\ 
14--15 & 271 & 0.0028 & 0.0052 \\ 
15--16 & 574 & 0.0029 & 0.0068 \\ 
16--17 & 1039 & 0.0030 & 0.0099 \\ 
17--18 & 2011 & 0.0031 & 0.0193 \\ 
18--19 & 3422 & 0.0035 & 0.0432 \\ 
19--20 & 4402 & 0.0048 & 0.0919 \\ 
20--21 & 5034 & 0.0107 & 0.2291 \\ 
21--23 & 326 & 0.0261 & 0.4431 \\ 
\hline
Total/Error & 17298 & 0.0062 & 0.1102 \\
\hline
\end{tabular}
\end{table}

\subsection{TESS Data}

The \textit{Transiting Exoplanet Survey Satellite} (TESS) has four 2K$\times$2K charge-coupled devices (CCDs). It covers a field of view (FOV) of 24$^{\circ}$~$\times$~96$^{\circ}$ and has an angular resolution of about 21 arcsec pixel$^{-1}$ \citep{ricker2015transiting}. TESS observes the sky in \textit{sectors} that last about 27 days, and the pointing shifts by 27$^{\circ}$ between sectors along the ecliptic. It provides time-series photometry in the 600--1000~nm range (roughly the Cousins $I$ band), which is useful for studies of stellar variability. Here we use \textit{TESS} full-frame image (FFI) data for the cluster field. The \textit{TESS} data used in this work are available through the Mikulski Archive for Space Telescopes (MAST): \href{http://dx.doi.org/10.17909/t9-nmc8-f686}{10.17909/t9-nmc8-f686}

\subsection{SPHEREx}

The SPHEREx mission performs an all-sky survey that delivers low-resolution spectra in the wavelength range 0.75–5 $\mu$m for each spatial resolution element on the sky ($\sim$6.2 arcsec per pixel) on a timescale of about two weeks \citep{crill2020spherex}. The spectral resolving power varies from approximately $R \sim 40$ in Band 1 (0.75–1.12 $\mu$m) to about $R \sim 130$ in Band 6 (4.41-5.01 $\mu$m). The instrument has a pixel scale of 6.2 arcsec per pixel. The first SPHEREx all-sky survey was completed in December 2025, and the corresponding Level 2 calibrated data products covering most of the sky were made publicly available roughly two months after the observations \citep{akeson2025spherex}.

For this study, we retrieved spectrophotometric measurements of two selected BSSs using the publicly available SPHEREx Spectrophotometry Tool. This tool derives photometry by performing forced PSF fitting on the Level-2-calibrated SPHEREx images.

\section{Cluster Membership Determination}
\label{sec:membership}

To select cluster members, we use a \textbf{GMM} based on Gaia astrometry. Berkeley~18 is at low Galactic latitude ($b = +5.1^{\circ}$), so many field stars are mixed with cluster stars. Consequently, simple cuts based on proper motion or parallax do not effectively separate members. The GMM models the observed astrometric distribution as a sum of Gaussian components (typically one for the cluster and one for the field) and assigns a membership probability to each star from its astrometric values. This approach works well when the cluster and field overlap in kinematics, and it has been used in recent open-cluster studies \citep{Gao2018, Agarwal2021, Qiu2024, belwal2026time}.

The input sample is constructed following \citet{Kuldeep2025}. We select Gaia~DR3 sources within a circular region around the cluster, retaining only those with reliable five-parameter astrometric solutions and valid photometric measurements. We apply quality cuts: positive parallaxes, proper-motion uncertainties $\leq \epsilon_{\mu}$, and \texttt{RUWE}~$\leq$~1.4 \citep{Lindegren2021}. To reduce field contamination, we first estimate the cluster center in ($\mu_{\alpha}^{*}$, $\mu_{\delta}$, $\varpi$) using the $k$-Nearest Neighbours (kNN) method \citep{Cover1967}. We then normalize the data and fit a two-component GMM using the Expectation-Maximization algorithm \citep{Dempster1977, McLachlan2000}. Each star receives a membership probability, and we keep stars with $p \geq p_\mathrm{min}$ as cluster members. The mean astrometric values are consistent with the literature \citep{Cantat-Gaudin20, Hunt2023}.

Using this method, we identify 798 probable members with $p>70\%$. We cross-match 693 of these stars with those in \citet{Hunt2023}, thereby supporting our selection. The mean proper motion from our member sample is $\mu_{\alpha}^{} = 0.78 \pm 0.15$~mas~yr$^{-1}$ and $\mu_{\delta} = -0.06 \pm 0.13$~mas~yr$^{-1}$. These values agree, within uncertainties, with those of \citet{Cantat-Gaudin20} and \citet{Hunt2023}.

\begin{figure*}
    \includegraphics[width=18cm,height=5.5cm]{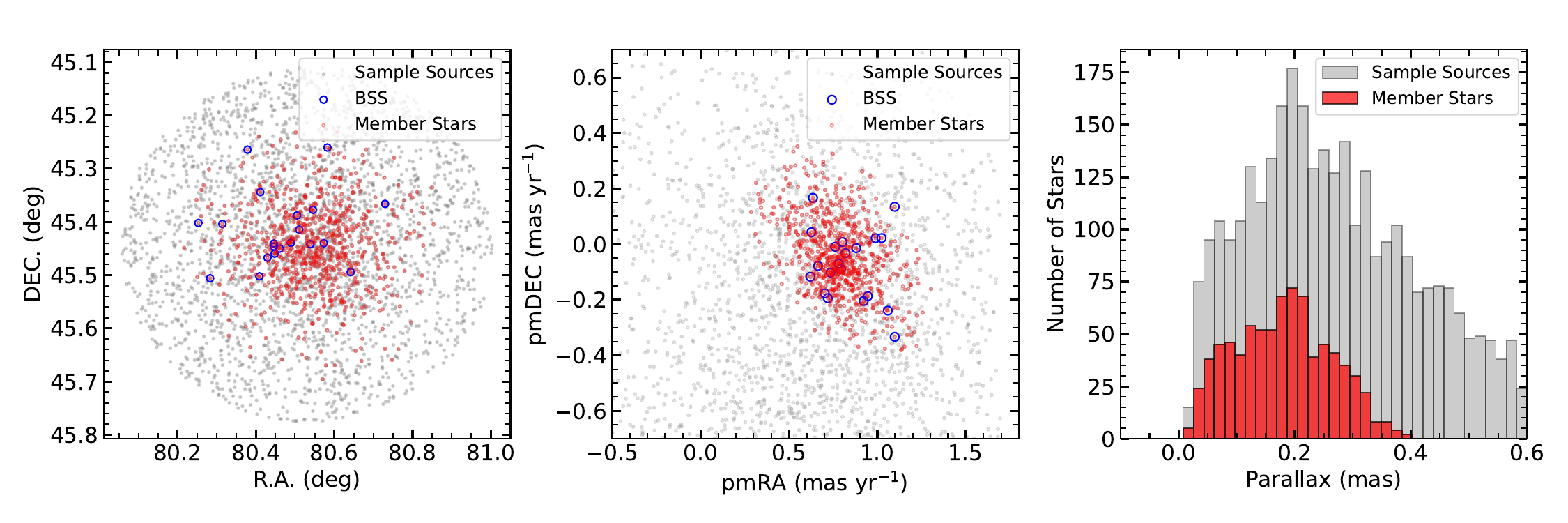}
    \hspace{-1.9cm}
    \caption{Spatial (left), proper-motion (middle), and parallax (right) distributions of stars in the field of Berkeley 18. Red points represent probable cluster members identified using the GMM analysis, blue circles denote BSSs, and grey points correspond to all sources in the field. The proper-motion and parallax panels demonstrate a clear concentration of cluster members, validating the adopted membership determination.}
    \label{members_both}
\end{figure*}

We check the sky positions, proper motions, and parallaxes of the selected members. In Figure~\ref{members_both}, member stars (red points) are more centrally concentrated on the sky than field stars (gray points). They also form a tight group on the proper-motion diagram and exhibit a narrow parallax peak relative to the field. This indicates that the GMM selection yields a relatively clean sample of members.

We estimate the distance using the mean parallax of the members. Since parallax errors are not always symmetric, we use the probabilistic distance method of \citet{bailer2018estimating}. The mean trigonometric parallax is $0.15 \pm 0.10$~mas, which gives a distance of $5.01^{+0.75}_{-0.55}$~kpc. This agrees with earlier results \citep[e.g.,][]{Cantat-Gaudin20, Hunt2023}.

\section{Structural Parameter Determination}
\label{sec:rdp}

To measure the internal structure of Berkeley~18, we constructed its radial density profile (RDP) using exclusively the high-probability cluster members identified via our Gaia astrometric analysis. We computed the surface density, $\rho(r_i)$, at radius $r_i$ by counting the number of stars ($N_i$) in concentric annuli and dividing by the annular area ($A_i$), i.e., $\rho(r_i)=N_i/A_i$. We then fit the observed profile using the empirical King surface-density model \citep{King62}, which describes how stellar density varies from the cluster center to the outer tidal limit. The model is
\begin{equation}
\rho(r) = \rho_0 \left[ \frac{1}{\sqrt{1 + (r/r_c)^2}} - \frac{1}{\sqrt{1 + (r_t/r_c)^2}} \right]^2 + \rho_{bg},
\label{eq:king_model}
\end{equation}
where $\rho_0$ is the central density, $r_c$ is the core radius (where $\rho = 0.5\,\rho_0$), $r_t$ is the tidal radius, and $\rho_{bg}$ is the background field density.

We estimate the best-fit parameters using a maximum-likelihood method. We minimize the negative log-likelihood
\begin{equation}
\ln \mathcal{L} = -\frac{1}{2}  \sum_i \left( \frac{\rho_i - \rho_{i,{\rm model}}}{\sigma_{\rho_i}} \right)^2,
\label{eq:log_likelihood}
\end{equation}
where $\rho_i$ is the observed density in each annulus, $\sigma_{\rho_i}$ is the Poisson uncertainty, and $\rho_{i,{\rm model}}$ is the model value.

\begin{figure*}
\centering
\includegraphics[width=0.48\linewidth]{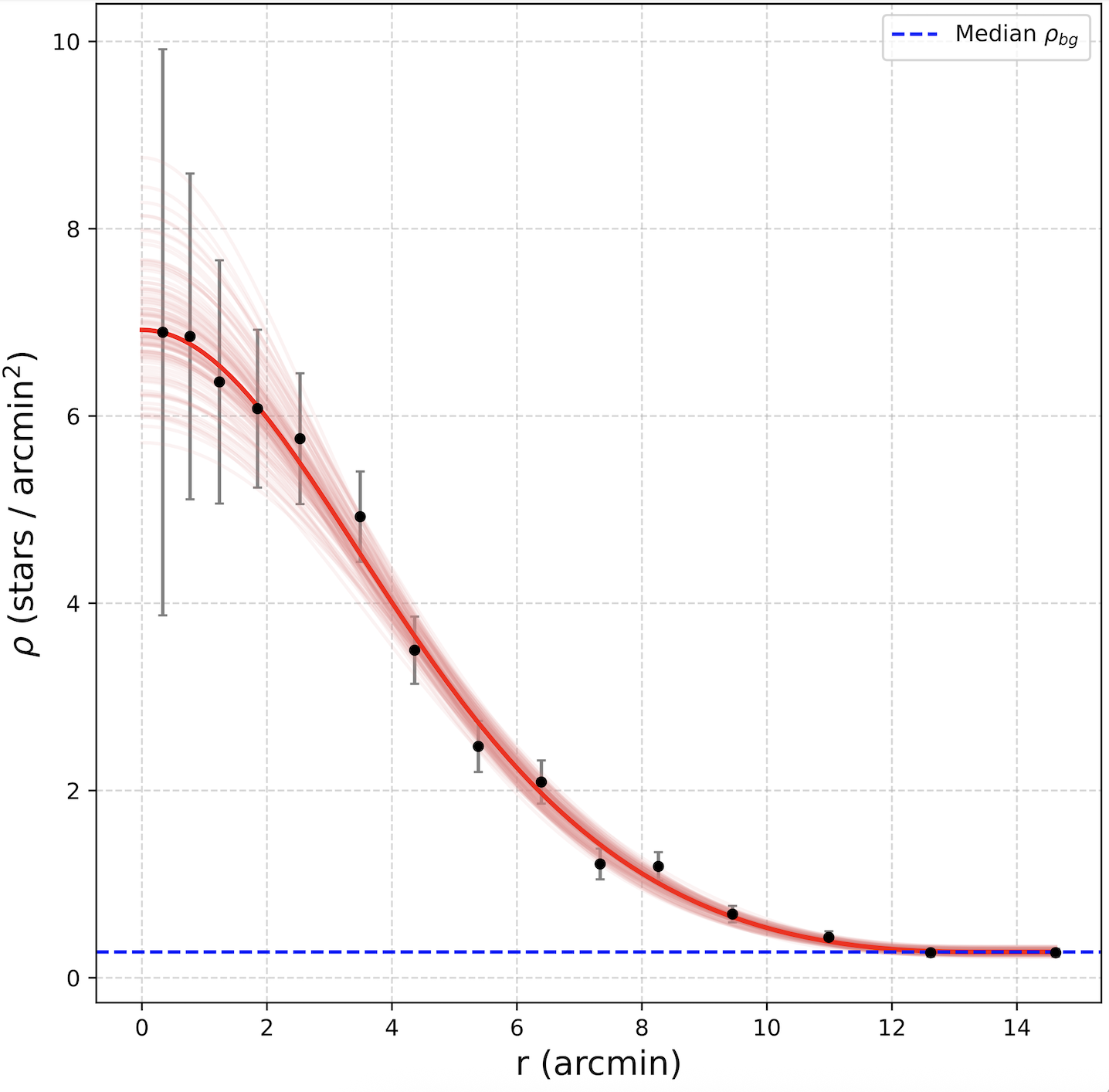}
\includegraphics[width=0.48\linewidth]{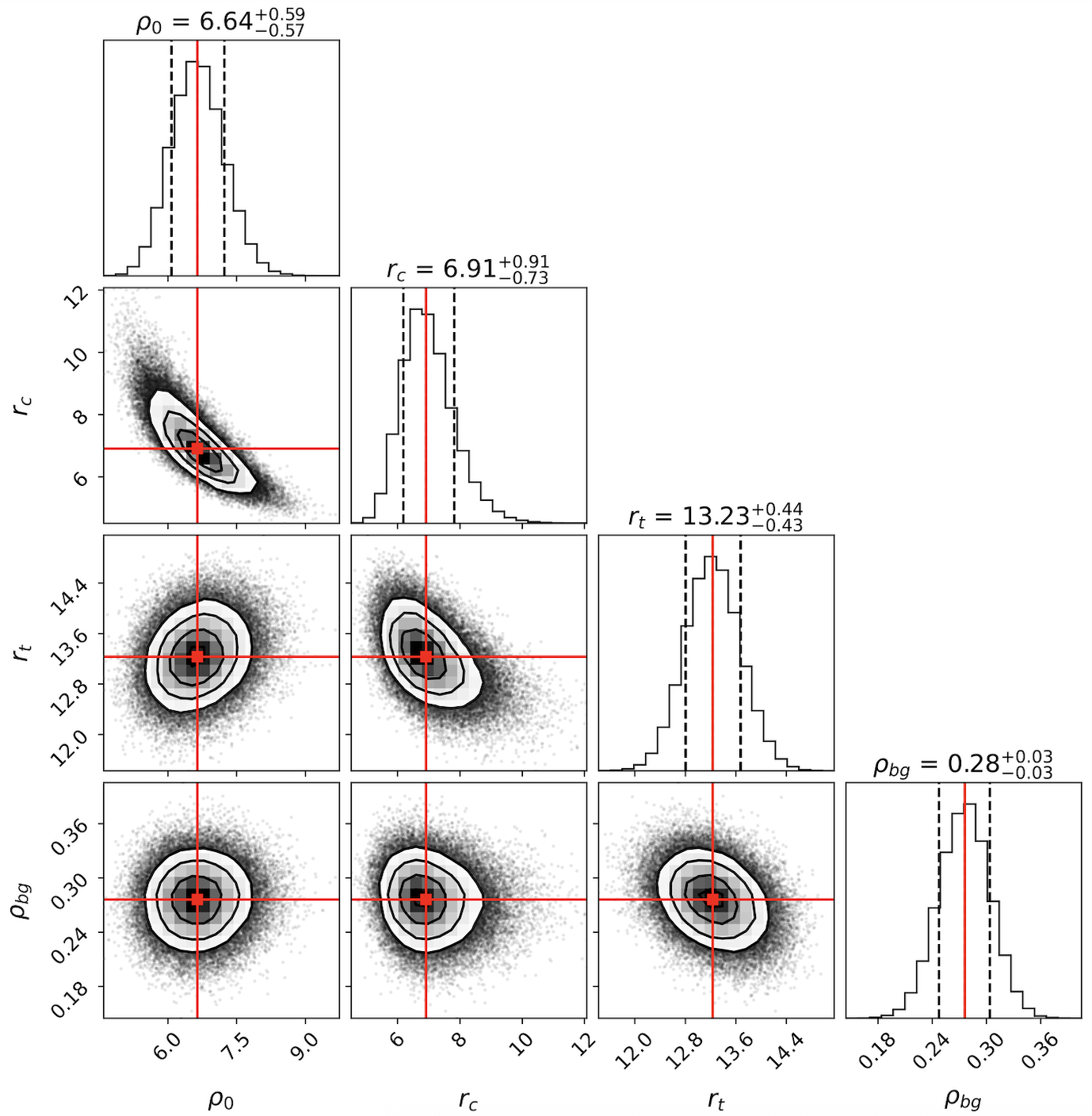}
\caption{Radial stellar density profile of the cluster. The left panel shows the observed stellar surface densities (black points with Poisson uncertainties) as a function of radial distance from the cluster center, along with the best-fitting  \citet{King62} model (solid red curve). The blue dashed line indicates the estimated background field density, and the shaded region represents the $1\sigma$ confidence interval of the model fit derived from the posterior distribution. The right panel shows the corresponding corner plot illustrating the posterior probability distributions and parameter correlations for the King-model parameters.}
\label{fig:RDP_fits}
\end{figure*}

We use the \texttt{emcee} sampler \citep{emcee} with 64 walkers and 7000 steps. We discard the first 1000 steps as burn-in. We assume uniform priors on all parameters. We check convergence using the Gelman--Rubin $\hat{R}$ statistic \citep{gelman1992} and find $\hat{R} \leq 1.2$. The quoted uncertainties on all structural parameters correspond to the 16th and 84th percentiles of the marginalised posterior distributions, equivalent to $1\sigma$ credible intervals. The best-fit profiles and posterior distributions are shown in Figure~\ref{fig:RDP_fits}, and the fitted parameters are listed in Table~\ref{tab:King_results}.

Following the classical definition of the King model \citep{King62}, the concentration parameter is defined as $C = \log(r_t/r_c)$. Using our derived core and tidal radii, we obtain a concentration parameter of $C = 1.91^{+0.30}_{-0.27}$, corresponding to a ratio of $r_t/r_c \sim 80$. This level of concentration is consistent with values reported for several dynamically evolved old open clusters \citep[e.g.,][]{rao2023determination}. For comparison, previous wide-field surveys have reported noticeably higher values for Berkeley~18, with \citet{Hunt2023} deriving $C = 3.49$ and \citet{Ahmed2026} reporting $C = 2.84$. This discrepancy is likely methodological: our RDP is constructed using only high-probability kinematic members, which significantly reduces background contamination. In contrast, earlier automated analyses often include all stars within the field of view, where field-star contamination can inflate the derived tidal radius and, consequently, the concentration parameter. This highlights the importance of robust membership selection in deriving reliable structural parameters for OCs \citep{Bilir2026}.

\begin{table}
\renewcommand{\arraystretch}{1.3}
\caption{Structural parameters of the cluster derived in this work from King-profile fitting to their radial density distributions. The table lists the central stellar density ($\rho_0$), background field density ($\rho_{\mathrm{bg}}$), core radius ($r_c$), tidal radius ($r_t$), and concentration parameter ($C$). All parameters are quoted with their $1\sigma$ uncertainties.}
\label{tab:King_results}
\centering
\setlength{\tabcolsep}{2pt}
\begin{tabular}{ccccc}\hline\hline
$\rho_0$ & $\rho_{bg}$ & $r_c$ & $r_t$ & $C$ \\
 \cline{1-2} \cline{3-4}
\multicolumn{2}{c}{(stars arcmin$^{-2}$)} & \multicolumn{2}{c}{(arcmin)} &  \\
\hline
 $6.64^{+0.59}_{-0.57}$ 
& $0.28^{+0.03}_{-0.03}$ 
& $6.91^{+0.91}_{-0.73}$ 
& $13.23^{+0.44}_{-0.43}$ 
& $1.91^{+0.30}_{-0.27}$  \\
\hline
\end{tabular}
\end{table}

\section{Color--magnitude diagrams and Distribution of BSS\lowercase{s}}\label{sec:CMD}

\begin{figure}
	\centering 
    \vspace{-0.1cm}
    \includegraphics[width=8.5cm,height=8.0cm]{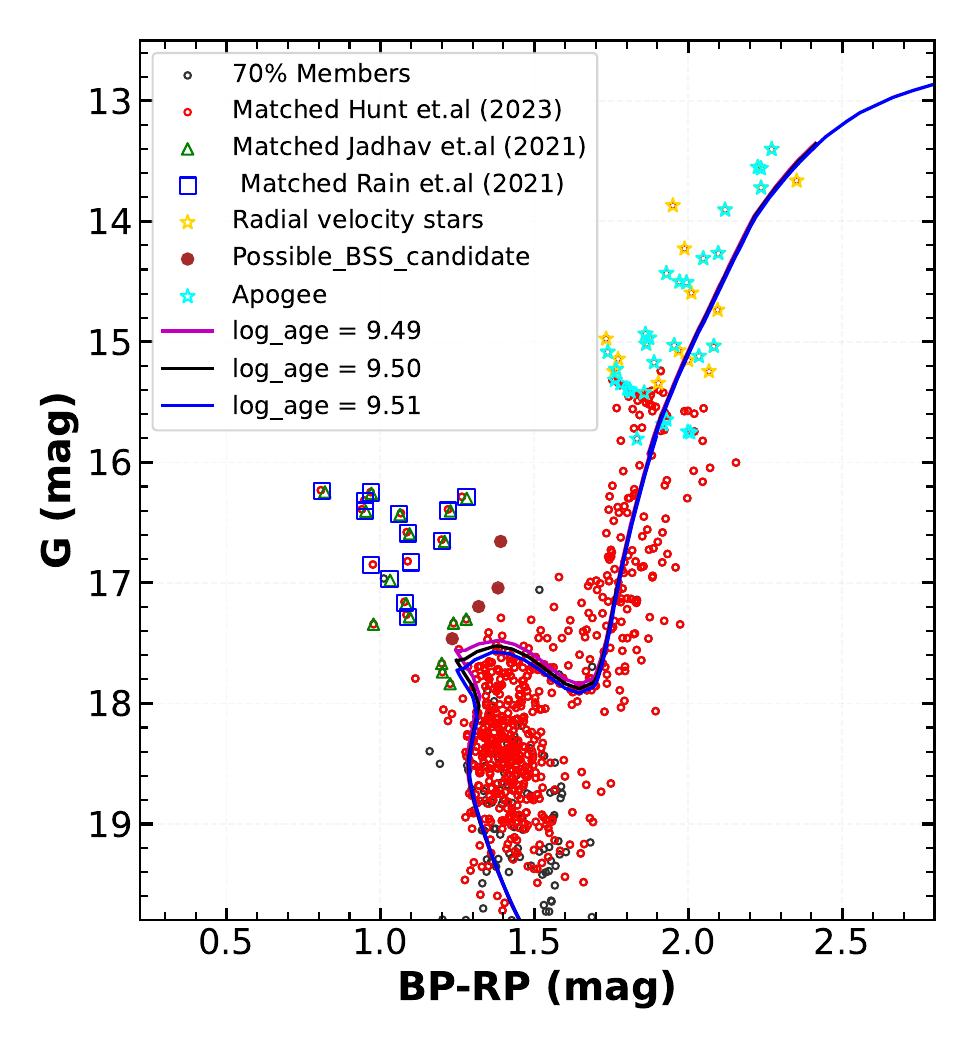}	
    \caption{CMD of the cluster, plotted in the $G$ versus $(\mathrm{BP}-\mathrm{RP})$ plane. Grey points represent all stars within the adopted cluster radius, while red points denote probable cluster members selected using astrometric criteria. The solid blue curves show the best-fitting PARSEC isochrones corresponding to the derived cluster parameters. For Berkeley~18, isochrones with metallicity(Z) = 0.006 $\pm$ 0.001 and $\log(\mathrm{age/yr}) = 9.49$, 9.50, and 9.51 are shown. Blue squares represent the BSS candidates reported by \citet{Rain2021}, while green triangles denote those identified by \citet{Jadhav2021a}. The filled circles mark the possible new BSS candidates identified in this study.}
	\label{fig:cmd}
\end{figure}

The photometric properties of Berkeley~18 are examined using the Gaia color--magnitude diagram (CMD), which provides a comprehensive view of the cluster's stellar population and evolutionary status. The morphology of the CMD enables the identification of key evolutionary features such as the main sequence, turn-off point, and red giant members, which are essential for deriving fundamental cluster parameters, including age, distance, and stellar masses \citep{bisht2019mass, bisht2020comprehensive, Bisht2026}. To ensure a clean representation of the cluster sequences, only stars with a membership probability $\geq 70\%$, as derived from the GMM analysis, are considered. The resulting CMD reveals well-defined evolutionary features, extending from the main sequence (MS) through the turn-off (TO) region to the red giant branch (RGB), consistent with the morphology expected for an old open cluster. Figure~\ref{fig:cmd} illustrates the CMD of Berkeley~18.

In our analysis, we estimated the cluster metallicity using the APOGEE spectroscopic dataset. We cross-matched our cluster members with the APOGEE catalogue and identified 33 stars in common, as shown in Figure~\ref{fig:cmd} (cyan star symbols). A Gaussian function was fitted to the [Fe/H] distribution to determine the mean metallicity, yielding [Fe/H] $= -0.378 \pm 0.04$ (corresponding to $Z = 0.006$). This value is in good agreement with that reported by \citet{netopil2016metallicity}, although their estimate was based on a single star. For the isochrone fitting, we explored models with different ages and found that isochrones with $\log(\mathrm{age/yr}) = 9.49$, 9.50, and 9.51 best represent the observed CMD, corresponding to an age of $3.2 \pm 0.2$~Gyr. The quoted uncertainty reflects the range of isochrone ages providing acceptable fits to the CMD when the distance modulus, reddening, and metallicity are each varied within their respective uncertainties; formal internal fitting errors alone yield $\pm 0.07$~Gyr, but we adopt the more conservative estimate to account for systematic effects.

From the isochrone fitting, we estimated a distance modulus of $15.1 \pm 0.20$ mag and a colour excess of $E(BP-RP) = 0.69 \pm 0.10$ mag. The uncertainties on these quantities were determined by varying each parameter independently about the best-fit value until the isochrone fit degraded noticeably, following the approach of \citet{bisht2019mass}. To evaluate the reliability of the fit, we applied the reduced chi-square test following the methodology described by \citep{valle2021goodness}. The best-fitting values of the distance modulus and colour excess were obtained by minimizing the $\chi^{2}$ statistic, which measures the deviation between the observed and model CMDs while accounting for observational uncertainties. The resulting reduced chi-square value of 0.80 indicates a satisfactory agreement between the model and the observations.

The MS is narrow and reaches a TO at $G \approx 17$~mag. We also observe a clear red clump, confirming the presence of an evolved population. PARSEC isochrones \citep{Bressan2012} generated using the CMD~3.9 interface\footnote{\url{http://stev.oapd.inaf.it/cgi-bin/cmd}} with $\log(\mathrm{age/yr}) \approx 9.50 \pm 0.01$ (corresponding to $\sim$3.2~Gyr) provide a reasonable match to the observed CMD, suggesting that Berkeley 18 belongs to the population of older OCs in the Galactic disk. We mark a few stars above the TO as possible BSS candidates. These stars are bluer and brighter than expected for the cluster TO and may have formed through binary mass transfer or stellar mergers. The number of identified BSS candidates (24) is consistent with expectations for old OCs of similar age and mass \citep[e.g.,][]{Ahumada2007, Rain2021}. This number is also consistent with expectations for dynamically evolved OCs in the outer Galactic disk, where binary evolution is expected to dominate over collisional processes.

We cross-matched our sample with the catalogues of \citet{Rain2021} and \citet{Jadhav2021a}, identifying 14 and 17 common stars, respectively, of which 11 stars are common to both catalogues.
We also overplot previously reported members  \citep[e.g.,][]{Hunt2023} and radial-velocity members from spectroscopic studies. Their agreement with our astrometric selection supports the reliability of the GMM membership probabilities. Overall, the CMD of Berkeley~18 shows features typical of an old open cluster at low Galactic latitude: clear evolutionary sequences and a likely BSS population. This makes Berkeley~18 a suitable target for studies of stellar evolution and cluster dynamics in the Galactic disk.

\begin{figure}
    \centering
    \includegraphics[width=\linewidth]{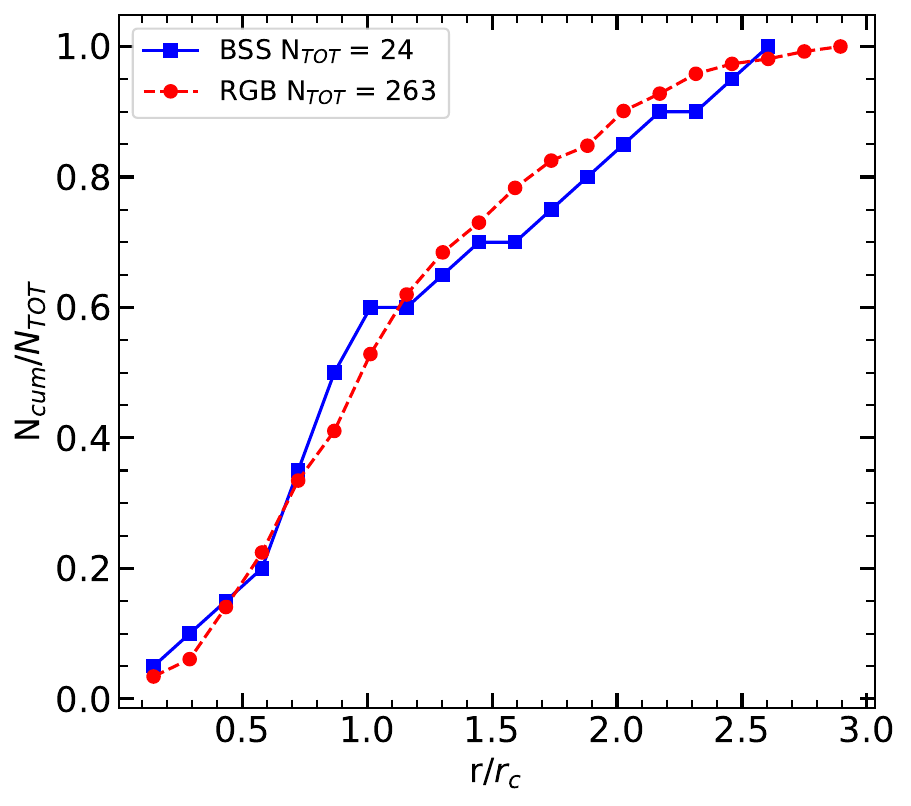}
    \caption{Cumulative radial distributions of BSSs (blue squares) and red giant branch stars (RGBs; red circles). The normalized cumulative fraction ($N_{\mathrm{cum}}/N_{\mathrm{tot}}$) is plotted as a function of projected radius in units of the cluster core radius ($r_c$). The vertical dashed line marks the radius within which BSSs are more centrally concentrated than RGB stars, beyond which their distributions converge. The Kolmogorov–Smirnov test yields $p \lesssim 0.03$.}
    \label{fig: radial_distribution}
\end{figure}

To examine the cumulative radial distributions, we adopted the red giant branch (RGB) stars as the reference population, following the method described by \citet{Vaidya2020}. Using this reference, we constructed the cumulative radial distributions for both the BSSs and RGB stars in the cluster. Figure~\ref{fig: radial_distribution} presents these distributions, where the normalized cumulative stellar counts are shown on the y-axis and the radial distance, expressed in units of the core radius ($r_c$), is plotted on the x-axis. To make a fair comparison between the two populations, we restricted the analysis to stars within the same magnitude range, as illustrated in Figure~\ref{fig: radial_distribution}.

Although the two distributions show a similar trend around 1.1$r_c$, the BSS population exhibits a mild central concentration relative to RGB stars in the inner regions; the overall distributions partially overlap, indicating incomplete dynamical segregation. In addition, the Kolmogorov–Smirnov test suggests that the BSS and RGB populations are unlikely to originate from the same parent distribution ($p \lesssim 0.03$), although some overlap is present in their radial distributions.

\subsection{A$^{+}$ Parameter Estimation}

The observational parameter A$^{+}$ serves as a measure of the degree of BSS segregation within the cluster core, systematically increasing as the host cluster evolves \citep{alessandrini2016investigating} .
To quantify the degree of mass segregation of BSSs in Berkeley 18, we computed the $A^{+}$ parameter, defined as the area between the cumulative radial distributions of the BSS population and a reference (REF) population. In this study, depending on the photometric completeness relative to the TO magnitude, we adopted subgiant branch (SGB), RGB, and red clump (RC) stars as the REF population.

The $A^{+}$ parameter is expressed as:
\begin{equation}
A^{+} = \int_{x_{\rm min}}^{x_{\rm max}} \left[ \phi_{\rm BSS}(x) - \phi_{\rm REF}(x) \right] dx,
\end{equation}
where $\phi_{\rm BSS}$ and $\phi_{\rm REF}$ are the cumulative radial distribution functions of the BSS and REF populations, respectively. The radial coordinate is defined as $x = \log (r/r_{h})$, where $r_{h}$ is the half-mass radius of the cluster.

Following \citet{lanzoni2016refining}, we computed $A^{+}$ within the half-mass radius ($r_{h}$), which provides a consistent and physically meaningful measure of BSS segregation and is particularly sensitive to dynamical friction effects in the cluster core \citep{rao2021determination}. The cumulative distributions were constructed as a function of the logarithmic radial distance normalized to $r_{h}$.

The projected half-light radius ($r_{hp}$) was estimated using the relation from \citet{santos2020viscacha}:
\begin{equation}
\log (r_{hp}/r_{c}) = -(0.339 \pm 0.009) + (0.602 \pm 0.015)C - (0.037 \pm 0.005)C^{2},
\end{equation}
where $C = \log (r_{t}/r_{c})$ is the concentration parameter. The half-mass radius was then derived using $r_{h} = 1.33\,r_{hp}$ \citep{baumgardt2010evidence}, yielding $r_{h} = 8.14$ arcmin.

\begin{figure}
    \centering
    \includegraphics[width=\linewidth]{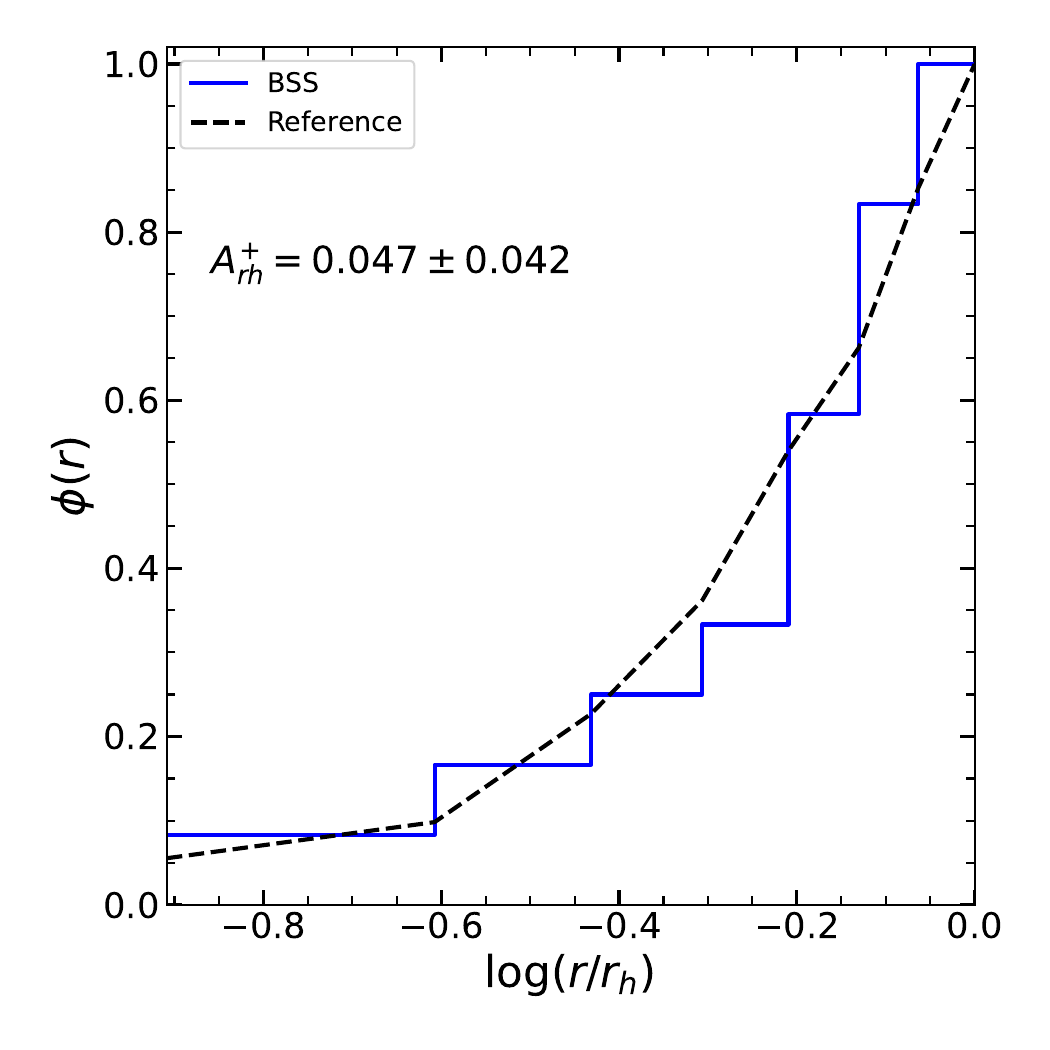}
    \caption{Cumulative radial distributions of the BSS population (blue solid line) and the reference (REF) population (black dashed line) as a function of logarithmic radial distance from the cluster centre, expressed in units of the half-mass radius ($r_h$). The area between the two distributions corresponds to the $A^{+}_{r_h}$ parameter, yielding $A^{+}_{r_h} = 0.047 \pm 0.042$. The small positive value indicates a mild central concentration of BSSs relative to the reference population, consistent with a dynamically intermediate state.}
    \label{fig: A_plus}
\end{figure}

The resulting cumulative radial distributions of the BSS and REF populations are shown in Figure~\ref{fig: A_plus}. The derived value of $A^{+}_{r_h} = 0.047 \pm 0.042$, which shows good agreement with \citet{rao2023determination}, indicates a mild degree of central concentration of BSSs relative to the reference population, consistent with a dynamically intermediate or weakly segregated state.

\section{Spectral Energy Distribution Analysis of Blue Straggler Stars}
\label{sec:sed}

We studied the spectral energy distributions (SEDs) of the BSS candidates in Berkeley~18 to estimate their fundamental stellar parameters, namely the effective temperature ($T_{\rm eff}$), radius ($R$), and luminosity ($L$). The SED fitting was performed using the Virtual Observatory SED Analyzer (VOSA; \citealt{Bayo2008}), combining optical (Gaia~DR3, Pan-STARRS, SkyMapper, TESS) and infrared (2MASS, WISE) photometry.

For each source, we adopted the cluster distance of $5.01^{+0.75}_{-0.55}$~kpc and a line-of-sight extinction of $1.64\pm0.17$ mag, derived from the isochrone analysis (Section~\ref{sec:CMD}). To ensure consistency, a uniform extinction law was applied using \citet{Fitzpatrick1999} in the optical and \citet{Indebetouw2005} in the infrared, assuming a standard value of $R_{\rm V} = 3.1$. We verified that allowing small variations in extinction ($\pm$ 0.1-0.2 mag) does not significantly affect the derived stellar parameters, indicating that the assumption of a uniform extinction is adequate for the present analysis. We note that adopting a single extinction value for all BSS candidates implicitly assumes negligible differential reddening across the cluster field. Given the relatively sparse nature of Berkeley~18 and the absence of strong spatial variations in the colour–magnitude diagram, this approach is expected to have a limited impact on the derived parameters. Photometric counterparts were retrieved within a $5$ arcsec radius. Each source was visually inspected using Aladin\footnote{\url{https://aladin.cds.unistra.fr}} to assess potential blending effects, and no obvious contamination was identified within the adopted matching radius, although unresolved blending cannot be ruled out.

The SEDs were fitted with single-star atmosphere models from the Kurucz ODFNEW/NOVER grid \citep{Castelli1997, Castelli2003}, covering $T_{\rm eff} = 6000$--$15000$~K and $\log g = 2.5$--$4.5$. To ensure a robust search for circumstellar infrared excess (Section \ref{sec:IR_excess}), the fits were anchored primarily to the optical photometry, where the BSS photospheric emission peaks, preventing potential infrared excess from biasing the derived stellar parameters. The best-fitting model was selected by minimizing the reduced chi-square statistic, $\chi^2_\nu$, between the observed and synthetic fluxes.

To improve the robustness against underestimated photometric uncertainties, we also used the VOSA goodness-of-fit parameter $V_{\rm gfb}$ \citep{Bayo2008}, defined as:
\begin{equation}
V_{\rm gfb}^2 = \frac{1}{N - M} \sum_{i=1}^{N} \left( \frac{F_{o,i} - m_d F_{m,i}}{a_{o,i}} \right)^2,
\end{equation}
where
\begin{equation}
a_{o,i} =
\begin{cases}
0.02,F_{o,i}, & \text{if } \sigma_{o,i} \leq 0.02,F_{o,i}, \\
\sigma_{o,i}, & \text{if } \sigma_{o,i} > 0.02,F_{o,i}.
\end{cases}
\end{equation}
Following standard VOSA practice, this definition effectively imposes a minimum relative uncertainty on flux. We also employed the \texttt{Binary\_SED\_Fitting} Python module \citep{2021JApA...42...89J, jadhav_2024_13928317}\footnote{\url{https://github.com/jikrant3/Binary_SED_Fitting}}. The resulting $\chi^2_\nu$ values and uncertainties are broadly consistent with those from the VOSA-based Monte Carlo analysis; for consistency across the sample, we adopt the VOSA results. We further apply a conservative acceptance threshold of $V_{\rm gfb} < 15$.

Most BSS candidates satisfy both $\chi^2_\nu$ and $V_{\rm gfb}$ criteria, indicating an overall reasonable agreement between the data and models. A subset shows relatively high $\chi^2_\nu$ values ($\chi^2_{\nu} \gtrsim 10$), which may be related to unresolved binarity, common among BSSs, since single-star models can yield elevated $\chi^2$ in the presence of additional flux from unseen companions \citep{Rani2021, 2022MNRAS.516.2444R, Sheikh2024}.

In evaluating fit quality, we therefore consider $V_{\rm gfb}$ alongside $\chi^2_\nu$. As shown in Table~\ref{tab:sed_parameters}, most sources have $V_{\rm gfb} < 1$, suggesting that the fits remain acceptable despite variations in $\chi^2$. No clear systematic residual trends with wavelength are found, supporting the consistency of the adopted fitting approach. Parameter uncertainties are estimated via Monte Carlo simulations within VOSA, using 100 iterations and propagating both photometric and distance errors. The resulting stellar parameters are listed in Table~\ref{tab:sed_parameters}.

\begin{figure}
    \centering
    \includegraphics[width=1\linewidth]{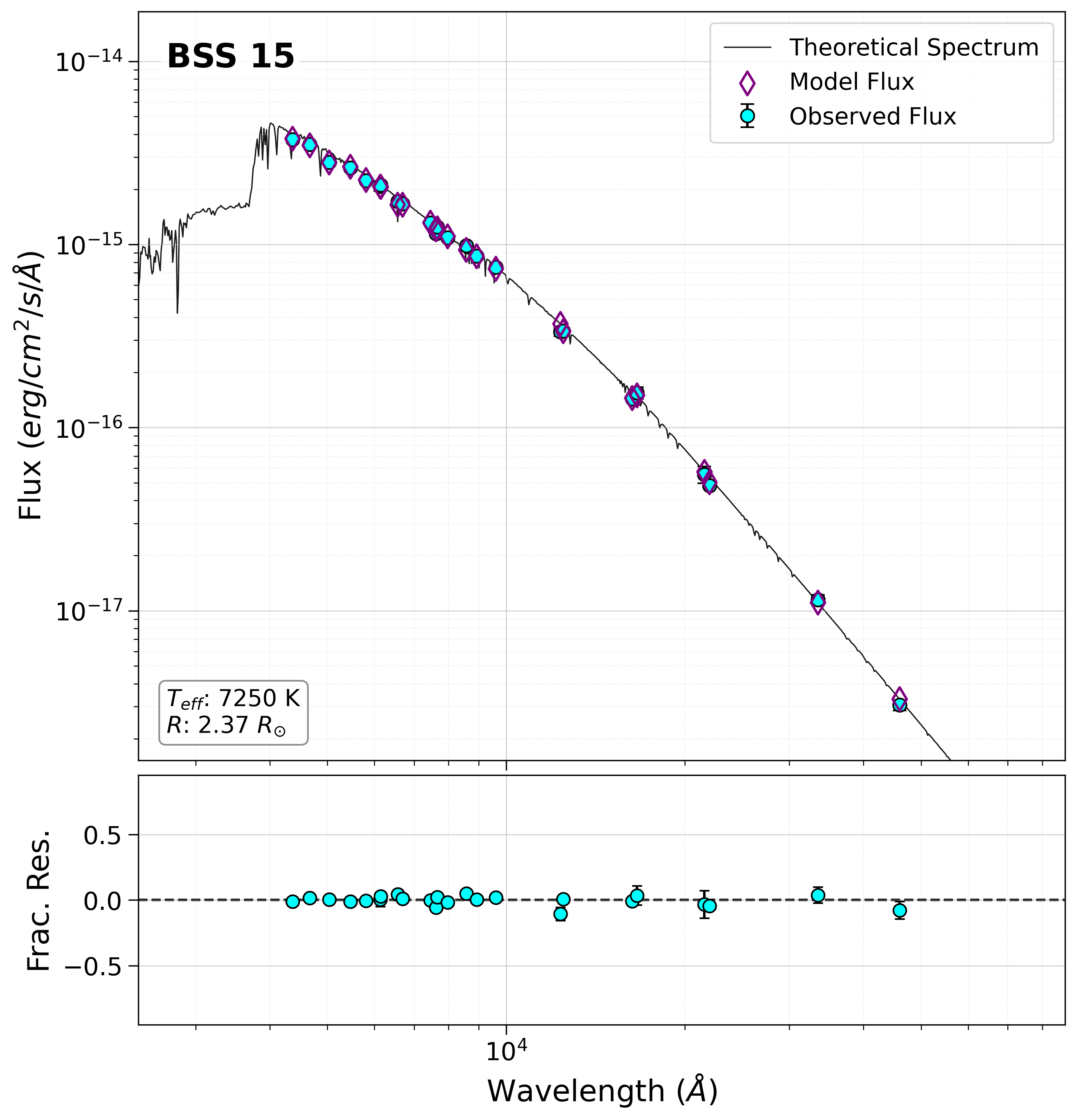}
    \caption{SED fitting for BSS 15. Top panel: Cyan circles with associated uncertainties denote the observed photometric fluxes, while the solid black curve represents the best-fitting \citet{Castelli2003} model spectrum. Purple diamonds indicate the synthetic fluxes computed for the corresponding photometric passbands. Bottom panel: Fractional residuals between the observed and model fluxes, illustrating the quality of the fit.}
    \label{fig:B18_SED}
\end{figure}

\begin{table*}
\centering
\caption{Stellar parameters derived from the SED fitting of the identified BSS candidates. The table provides the Gaia DR3 source identifiers, equatorial coordinates (J2000), effective temperatures ($T_{\rm eff}$), surface gravities ($\log g$), stellar radii ($R$), luminosities ($L$), SED scaling factors ($M_{\rm d}$), the number of photometric data points used in the fitting ($N_{\rm fit}$), the reduced chi-square ($\chi^2_\nu$), and the goodness-of-fit parameter ($V_{\rm gfb}$). All uncertainties correspond to 1 $\sigma$ confidence intervals.}
\resizebox{\linewidth}{!}{
\begin{tabular}{ccllcccccccr}
\hline
\hline
BSS ID & $Gaia$ DR3 Source ID & $\alpha$ (J2000) & $\delta$ (J2000) & $T_{\rm eff}$ (K) & $\log g$ & $R$ ($R_\odot$) & $L$ ($L_\odot$) & $M_{\rm d}$ & $N_{\rm fit}$ & $\chi^2_\nu$ & $V_{\rm gfb}$ \\\hline
BSS 01 & 208544267362577536 & 05:21:30.87 & +45:15:51.23 & 6250$\pm$224 & 3.50$\pm$0.37 & 1.54$\pm$0.14 & 3.26$\pm$1.02 & 6.63E-23 & 12 & 1.47 & 0.34 \\
BSS 02 & 208546535104330752 & 05:22:11.01 & +45:22:38.85 & 6750$\pm$228 & 3.00$\pm$0.40 & 2.95$\pm$0.26 & 16.23$\pm$4.84 & 2.12E-22 & 21 & 2.02 & 0.08 \\
BSS 03 & 208547737695033728 & 05:22:19.79 & +45:15:36.86 & 6500$\pm$203 & 3.00$\pm$0.44 & 2.03$\pm$0.18 & 6.69$\pm$1.91 & 6.82E-23 & 11 & 1.20 & 0.21 \\
BSS 04 & 208548665408081920 & 05:22:55.15 & +45:21:57.88 & 7750$\pm$204 & 3.50$\pm$0.25 & 1.46$\pm$0.14 & 6.93$\pm$1.92 & 8.01E-23 & 21 & 10.07 & 3.41 \\
BSS 05 & 208557873819068672 & 05:21:38.69 & +45:20:38.72 & 7250$\pm$198 & 3.00$\pm$0.38 & 2.62$\pm$0.26 & 17.15$\pm$5.02 & 1.38E-22 & 24 & 1.96 & 0.39 \\
BSS 06 & 208558251775703552 & 05:22:02.70 & +45:22:56.99 & 6500$\pm$125 & 4.00$\pm$0.45 & 4.14$\pm$0.41 & 27.53$\pm$7.07 & 2.42E-22 & 22 & 10.73 & 0.44 \\
BSS 07 & 208558320495187200 & 05:22:01.20 & +45:23:16.06 & 6750$\pm$250 & 3.00$\pm$0.30 & 3.86$\pm$0.35 & 28.21$\pm$9 & 1.26E-22 & 25 & 6.60 & 0.59 \\
BSS 08 & 208558599670803456 & 05:22:01.35 & +45:24:52.43 & 6750$\pm$173 & 3.00$\pm$0.42 & 2.40$\pm$0.23 & 11.15$\pm$3.06 & 8.90E-23 & 16 & 7.89 & 2.61 \\
BSS 09 & 208558767172290048 & 05:21:33.81 & +45:22:02.78 & 7000$\pm$125 & 3.00$\pm$0.34 & 2.03$\pm$0.20 & 8.88$\pm$2.15 & 8.23E-23 & 21 & 5.17 & 0.24 \\
BSS 10 & 208559458664247680 & 05:22:03.67 & +45:26:49.85 & 6750$\pm$141 & 4.00$\pm$0.33 & 2.58$\pm$0.26 & 12.32$\pm$3.28 & 1.35E-22 & 21 & 8.22 & 1.25 \\
BSS 11 & 208559523086067840 & 05:21:57.40 & +45:26:21.85 & 6750$\pm$198 & 3.00$\pm$0.35 & 2.24$\pm$0.22 & 9.48$\pm$2.85 & 1.01E-22 & 18 & 8.48 & 0.55 \\
BSS 12 & 208559561743484544 & 05:21:46.99 & +45:26:26.32 & 7250$\pm$209 & 3.00$\pm$0.25 & 2.86$\pm$0.29 & 20.48$\pm$6.4 & 1.73E-22 & 24 & 3.01 & 0.44 \\
BSS 13 & 208559729244528384 & 05:22:04.55 & +45:28:44.05 & 6500$\pm$125 & 3.50$\pm$0.29 & 3.45$\pm$0.34 & 19.15$\pm$4.78 & 1.63E-22 & 20 & 11.19 & 0.29 \\
BSS 14 & 208560729974633344 & 05:21:15.45 & +45:24:13.95 & 6500$\pm$285 & 3.00$\pm$0.35 & 4.83$\pm$0.45 & 37.32$\pm$11.78 & 2.91E-22 & 19 & 17.07 & 0.19 \\
BSS 15 & 208561103634525312 & 05:21:00.72 & +45:24:07.26 & 7250$\pm$196 & 3.00$\pm$0.27 & 2.37$\pm$0.24 & 13.92$\pm$4.15 & 1.14E-22 & 24 & 11.81 & 0.16 \\
BSS 16 & 208562413601763456 & 05:21:43.06 & +45:28:02.72 & 7750$\pm$262 & 3.00$\pm$0.25 & 1.54$\pm$0.14 & 7.7$\pm$2.34 & 8.44E-23 & 24 & 2.09 & 3.02 \\
BSS 17 & 208562516680976896 & 05:21:50.59 & +45:27:00.59 & 7500$\pm$191 & 3.00$\pm$0.25 & 2.91$\pm$0.28 & 24.18$\pm$6.81 & 1.63E-22 & 20 & 2.10 & 0.19 \\
BSS 18 & 208562516680982656 & 05:21:47.13 & +45:26:46.73 & 6000$\pm$227 & 4.00$\pm$0.25 & 5.67$\pm$0.57 & 37.95$\pm$11.76 & 6.52E-22 & 23 & 11.08 & 0.75 \\
BSS 19 & 208562585400453120 & 05:21:47.47 & +45:27:35.02 & 7000$\pm$136 & 3.00$\pm$0.29 & 2.82$\pm$0.28 & 17.21$\pm$4.53 & 1.61E-22 & 24 & 2.38 & 0.49 \\
BSS 20 & 208563508816270464 & 05:21:38.13 & +45:30:08.39 & 7500$\pm$276 & 3.00$\pm$0.30 & 2$\pm$0.18 & 11.44$\pm$3.58 & 8.49E-23 & 24 & 16.85 & 0.66 \\
BSS 21 & 208564475186034048 & 05:22:17.62 & +45:26:23.92 & 8250$\pm$133 & 3.00$\pm$0.25 & 2.33$\pm$0.23 & 22.52$\pm$5.65 & 1.34E-22 & 21 & 2.11 & 0.68 \\
BSS 22 & 208565059301539328 & 05:22:34.10 & +45:29:39.80 & 7000$\pm$297 & 3.00$\pm$0.30 & 1.40$\pm$0.14 & 4.26$\pm$1.49 & 7.62E-23 & 21 & 3.92 & 1.42 \\
BSS 23 & 208565261161830784 & 05:22:09.34 & +45:26:30.69 & 7000$\pm$159 & 3.50$\pm$0.30 & 2.01$\pm$0.20 & 8.69$\pm$2.38 & 1.01E-22 & 21 & 6.43 & 0.52 \\
BSS 24 & 208749807317373568 & 05:21:07.94 & +45:30:21.26 & 8500$\pm$176 & 3.50$\pm$0.25 & 2.55$\pm$0.25 & 30.67$\pm$7.75 & 1.23E-22 & 24 & 6.07 & 0.55 \\
\hline
\end{tabular}
}
\label{tab:sed_parameters}
\end{table*}

Recent studies demonstrate that combining Gaia DR3 membership kinematics with wide-wavelength SED fitting provides a highly effective framework for constraining the fundamental parameters of BSSs \citep{Sheikh2024,Zeng2025}. The resulting SED fits for our targets are presented in Figure~\ref{fig:B18_SED}, with the complete set provided in Appendix~\ref{fig:appendix_sed_1} and Appendix~\ref{fig:appendix_sed_2}.

The physical parameters derived from our analysis reveal a diverse BSS population in Berkeley~18, spanning effective temperatures of $T_{\rm eff} \approx 6000$--$8500$~K. The inferred stellar radii extend from relatively compact sizes ($R \approx 1.4\,R_{\odot}$) up to significantly inflated structures ($R \approx 5.7\,R_{\odot}$), while luminosities range from $L \approx 3.3$ to $38\,L_{\odot}$. This broad parameter space translates directly to a wide variety of evolutionary states within the cluster.

\begin{figure}
    \centering
    \includegraphics[width=1\linewidth]{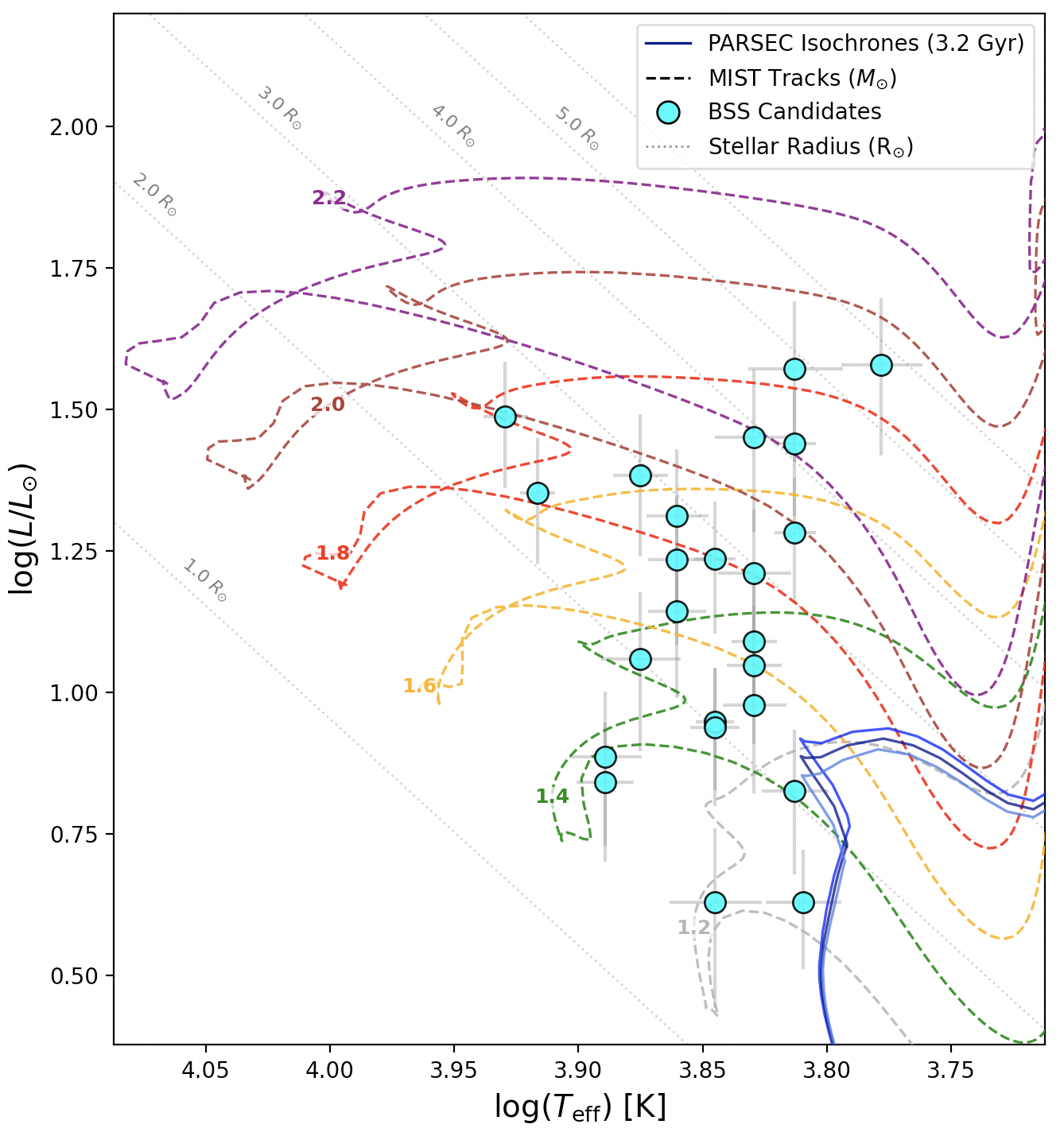}
    \caption{HR diagram of the cluster BSS candidates. Cyan circles show observed stellar parameters with error bars. Solid blue lines are PARSEC isochrones \citep{Bressan2012} for $\log(\mathrm{Age}) = 9.49$, 9.50, and 9.51. Colored dashed lines show MIST single-star evolutionary tracks for initial masses $1.2$–$2.2\,M_\odot$ \citep{MIST}, labeled in the figure. Grey dotted lines indicate constant stellar radii from 1.0 to 6.0\,$R_\odot$.}
    \label{fig:HR_SED}
\end{figure}

To assess the evolutionary status of the BSS candidates, we compared their positions in the Hertzsprung–Russell (HR) diagram (Figure~\ref{fig:HR_SED}) with the MESA Isochrones and Stellar Tracks (MIST)\footnote{\url{https://waps.cfa.harvard.edu/MIST/model_grids}} \citep{MIST}. Since BSSs formed via mass transfer or mergers may acquire angular momentum, we used rotating models with $v/v_{\rm crit} = 0.4$ and scaled them to the cluster metallicity. While the canonical MSTO of Berkeley~18 corresponds to about $1.2\,M_{\odot}$, many candidates lie above this limit, extending toward the subgiant branch. The most luminous and expanded stars, such as BSS~14 and BSS~18 ($R \ge 4.8\,R_{\odot}$, $L \ge 37\,L_{\odot}$), occupy positions consistent with the subgiant phases of $1.8$-$2.0\,M_{\odot}$ tracks. This is consistent with scenarios in which these stars may have undergone significant mass accretion and are evolving away from the main sequence. In contrast, more compact candidates ($R \le 2\,R_{\odot}$) align with $1.4$-$1.6\,M_{\odot}$ tracks, which may be consistent with lower mass-transfer efficiency, different progenitor mass ratios, or more recent interaction events.

Ultimately, while single-star SED fitting provides robust constraints on the present-day photospheric properties of these rejuvenated stars, it cannot independently confirm their binary origins or interaction history. Nevertheless, the derived stellar parameters reveal a heterogeneous BSS population spanning a wide range of effective temperatures, radii, and luminosities, indicating evolutionary stages ranging from near-main-sequence objects to more evolved subgiant-like stars. This diversity is consistent with expectations from binary evolution scenarios, in which mass transfer or merger events at different epochs can produce BSSs with a range of masses and evolutionary states. Establishing a mass-transfer formation scenario, therefore, requires additional observational diagnostics sensitive to past interactions. In the following section, we examine infrared excess as an independent tracer of circumbinary material potentially associated with previous mass-transfer episodes. We emphasize that complementary approaches, such as spectroscopy, radial-velocity monitoring, and high-precision variability studies, are essential for directly confirming binarity and constraining the evolutionary pathways of BSSs.

\section{The Impact of the Dynamical Environment on Determining BSS Origin}
\label{sec:bss_origin}

The dynamical environment of a cluster not only shapes its overall structure but also influences the relative importance of different BSS formation mechanisms. Clusters experiencing strong tidal forces, such as those passing near the Galactic center or following highly eccentric orbits, may lose stars more efficiently or experience enhanced stellar interactions, potentially favoring collision-induced BSS formation \citep[e.g.,][]{2024A&A...686A.215L}. Conversely, clusters in the outer Galactic disk with more circular orbits are subject to weaker tidal perturbations, allowing binary systems to evolve relatively undisturbed through mass transfer or mergers \citep[e.g.,][]{2024ApJ...961..251Y}. Understanding these environmental effects is crucial, as they can leave observable signatures in the properties of BSSs, including infrared excesses due to circumstellar dust and correlations with orbital parameters. In the following sections, we explore these indicators in detail, examining both the potential for circumstellar material around BSS candidates and the orbital characteristics of their host clusters.

To evaluate the role of dynamical interactions in the formation of BSSs, we estimated the stellar collision-rate proxy, $\Gamma \propto \rho_c^2 r_c^3 / \sigma$ (e.g., \citet{davies2004}; \citet{liegh2007}), where $\rho_c$, $r_c$, and $\sigma$ represent the central stellar density, core radius, and velocity dispersion, respectively. Using the structural parameters derived in this work ($\rho_c = 6.64~\mathrm{stars~arcmin^{-2}}$, $r_c = 6.91$ arcmin), and adopting a typical open cluster velocity dispersion of $\sigma \sim 1~\mathrm{km~s^{-1}}$, we obtain a very low collision-rate proxy (of order $\Gamma \sim 10^{2}$ in relative units). This value is several orders of magnitude smaller than those of dense globular clusters ($\Gamma \sim 10^{5}$--$10^{7}$), indicating that direct stellar collisions are highly unlikely to contribute significantly to the BSS population. This extremely low value of the collision-rate proxy strongly disfavors a predominantly collisional origin for the BSS population in Berkeley~18 and instead supports binary evolution as the dominant formation channel, although a minor contribution from dynamical interactions cannot be entirely excluded.

\subsection{Assessment of Mid-Infrared Excess}\label{sec:IR_excess}

Infrared excesses based on WISE W3/W4 photometry have been widely used to identify circumstellar dust and debris disk candidates in stellar systems \citep{patel14}. However, such detections are highly susceptible to background contamination and blending, particularly in crowded cluster environments, and therefore require careful validation. In this study, we initially used VOSA \citep{Bayo2008} to search for circumstellar material around the identified BSS candidates. Automated broadband SED fitting suggested a possible mid-infrared (IR) excess in two sources, BSS~11 and BSS~17. In both cases, the catalogued mid-IR fluxes appear systematically higher than those predicted by the best-fitting single-star atmospheric models.

\begin{figure}
    \centering
    \includegraphics[width=1\linewidth]{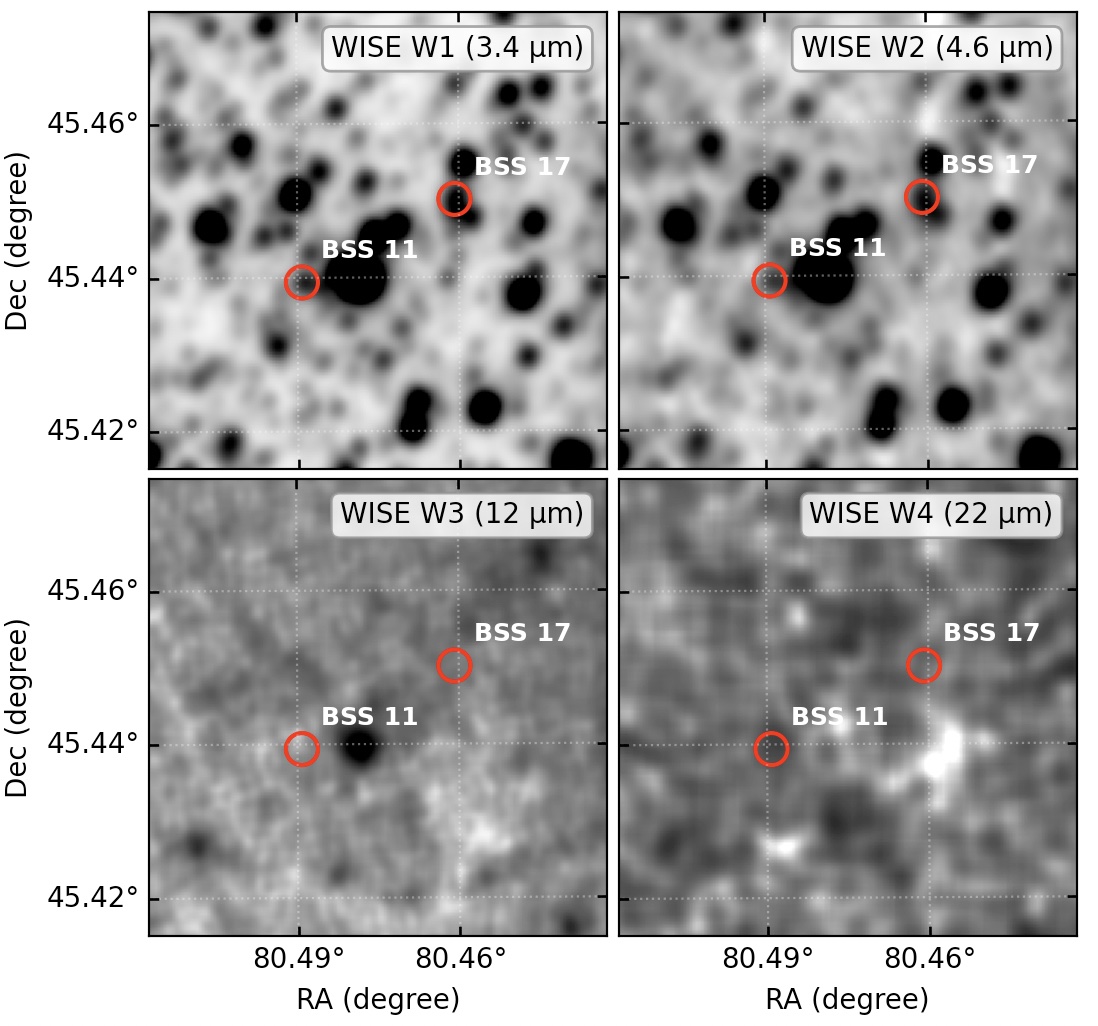}
    \caption{WISE multi-band (W1–W4) images of the regions surrounding BSS 11 and BSS 17. The red circles mark the astrometric positions of the candidates. In the longer-wavelength bands (W3 and W4), BSS 11 appears to be significantly blended with nearby sources, likely contaminating its mid-infrared photometry. In contrast, BSS 17 is located in a relatively isolated region and shows no clear evidence of a genuine mid-infrared excess. This comparison highlights the importance of visual inspection in assessing the reliability of infrared excess detections.}
    \label{fig:wise_comparison}
\end{figure}

To check whether the catalogued mid-infrared excesses are real, we visually inspected the WISE \citep{Wright2010} images using the NASA/IPAC Infrared Science Archive (IRSA) WISE Image Service\footnote{\url{https://irsa.ipac.caltech.edu/applications/wise}}. As shown in Figure~\ref{fig:wise_comparison}, the apparent excess in BSS~11 is likely unreliable due to blending with a nearby bright source, which dominates the W3 and W4 bands. BSS~17, on the other hand, is in a relatively clean field and detected in W1 and W2, but shows no obvious emission in W3 or W4 at the stellar position. This suggests that the catalogued excess for BSS~17 may be spurious rather than indicative of real circumstellar emission.

To further check these results, we used independent low-resolution spectrophotometry from the SPHEREx mission \citep{Dore2014}, which provides continuous coverage from 1 to 5~$\mu$m. The data were filtered (quality flag = 2097152) to remove anomalies such as saturation or bad pixels. Comparing the SPHEREx spectra with our models, both BSS~11 and BSS~17 follow the expected Rayleigh-Jeans tail of the stars up to 5~$\mu$m (see Figure~\ref{fig:IR_Excess}). This indicates no significant near-IR excess, limiting the presence of hot circumstellar dust.

Overall, combining visual inspection of WISE images with SPHEREx spectrophotometry suggests that the catalogued mid-IR excesses for BSS~11 and BSS~17 are likely due to background contamination, source blending, or catalog issues. These results highlight the importance of careful checks when using automated photometric catalogs to identify circumstellar material in crowded cluster fields. The absence of confirmed mid-infrared excess among the BSS candidates suggests that there is no strong evidence for the presence of circumstellar dust associated with recent mass-transfer events. This finding is consistent with the expectation that any such material, if produced during earlier evolutionary phases, has either dissipated or is below the current detection limits.

\subsection{Galactic Orbital Context for BSS Formation}

In this study, we computed the present-day position and space velocity of Berkeley~18. We calculated the velocity components ($U$, $V$, $W$) in a right-handed Galactic coordinate system. The uncorrected values are ($U$, $V$, $W$) = (1.24~$\pm$~8.26, $-12.46$~$\pm$~0.35, 14.15~$\pm$~8.61)~km~s$^{-1}$. We corrected for the Solar motion relative to the Local Standard of Rest (LSR) using \citet{Coskunoglu2011}: ($U$, $V$, $W$)$_{\rm LSR}$ = (8.83~$\pm$~0.24, 14.19~$\pm$~0.34, 6.57~$\pm$~0.21)~km~s$^{-1}$. The resulting LSR-corrected velocities and total space velocity ($S_{\rm LSR}$) are given in Table~\ref{Tab:dyn_params_new}.

\begin{table*}
\centering
\footnotesize
\setlength{\tabcolsep}{1.1pt} 
\caption{Derived kinematic and orbital parameters of the open cluster Berkeley 18. The table lists the Galactocentric distance ($R_{\mathrm{GC}}$), velocity components relative to the local standard of rest ($U_{\mathrm{LSR}}, V_{\mathrm{LSR}}, W_{\mathrm{LSR}}$), total space velocity ($S_{\mathrm{LSR}}$), maximum vertical height ($Z_{\max}$), apogalactic distance ($R_a$), perigalactic distance ($R_p$), mean Galactocentric radius ($R_m$), orbital eccentricity ($e$), and orbital period ($T_p$).}
\label{Tab:dyn_params_new}
\begin{tabular}{ccccccccccc}
\hline
$R_{\rm gc}$ & $U_{\rm LSR}$ & $V_{\rm LSR}$ & $W_{\rm LSR}$ & $S_{\rm LSR}$ 
& $Z_{\rm max}$ & $R_{\rm a}$ & $R_{\rm p}$ & $R_{\rm m}$ & $e$ & $T_{\rm p}$ \\
(kpc) & ($\mathrm{km\,s^{-1}}$) & ($\mathrm{km\,s^{-1}}$) & ($\mathrm{km\,s^{-1}}$) & ($\mathrm{km\,s^{-1}}$)
& (kpc) & (kpc) & (kpc) & (kpc) &  & (Myr) \\
\hline
$12.86 \pm 1.08$ 
& $10.07 \pm 8.27$ 
& $1.73 \pm 0.49$ 
& $20.72 \pm 8.61$ 
& $23.10 \pm 11.95$ 
& $1.02 \pm 0.58$ 
& $15.04 \pm 1.98$ 
& $12.71 \pm 0.84$ 
& $13.87 \pm 1.41$ 
& $0.08 \pm 0.03$ 
& $412 \pm 47$ \\
\hline
\end{tabular}
\end{table*}

To compute the cluster orbit, we used the six-dimensional (6D) phase-space data from Section~\ref{sec:CMD}: $(\alpha,\delta)$, distance $d$, proper motions $(\mu_\alpha \cos\delta, \mu_\delta)$, and radial velocity ($V_R$). The orbit was integrated with the \texttt{galpy} code \citep{Bovy2015} using the axisymmetric \texttt{MWPotential2014} Galactic potential \citep{Cinar2024, Cinar2025, Cinar2026}.

\begin{figure}
\centering
\includegraphics[width=0.9\linewidth]{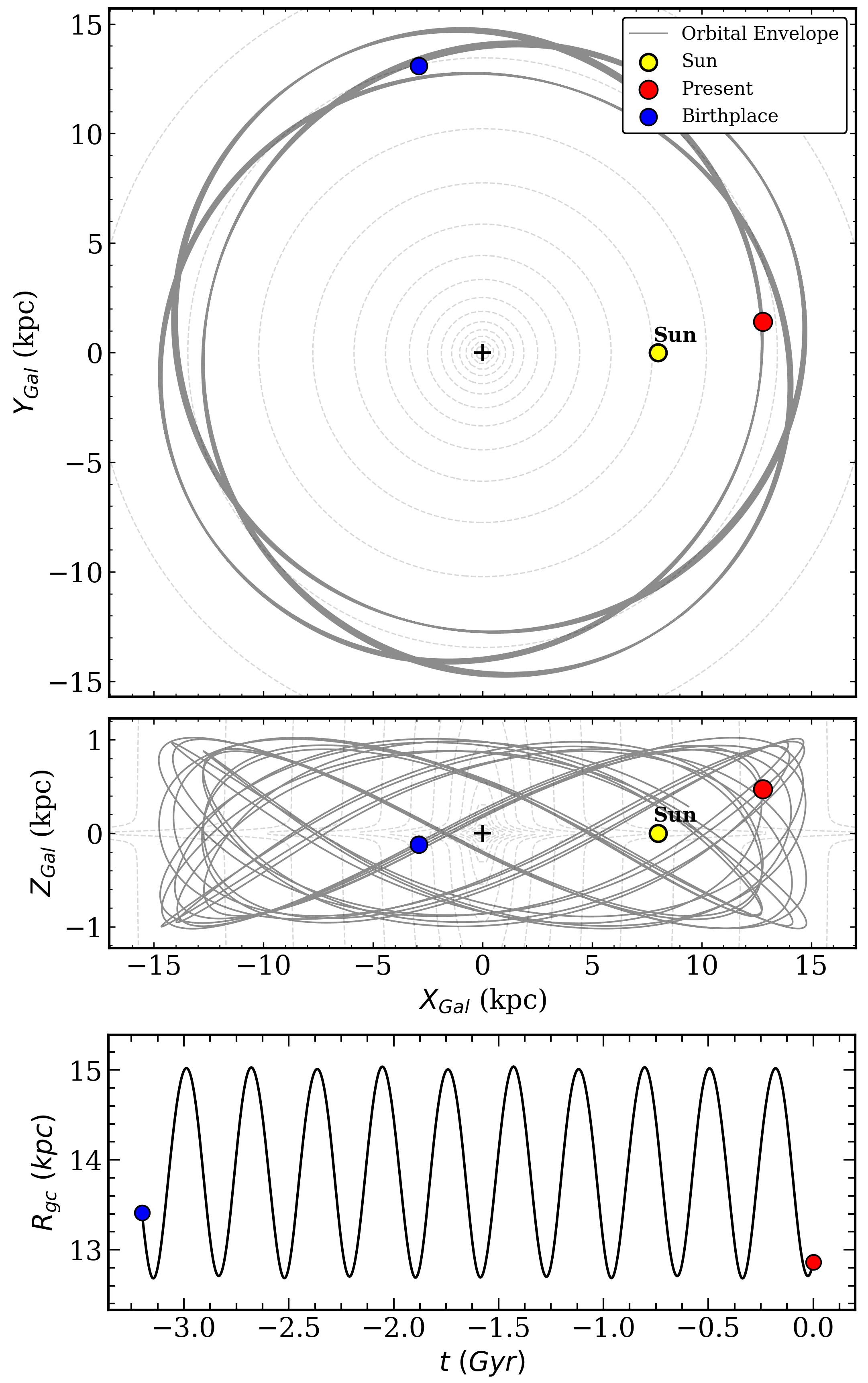}
\caption{Galactic orbital projections and radial evolution of Berkeley 18. \textbf{Top panel:} The cluster's orbit projected onto the Galactic $X$--$Y$ plane. \textbf{Middle panel:} The orbit projected onto the Galactic $X$--$Z$ plane. In the top and middle panels, the grey curves represent the integrated orbital envelope. The yellow, red, and blue circles indicate the position of the Sun, the present-day position of the cluster, and its inferred birthplace, respectively. \textbf{Bottom panel:} The galactocentric radius ($R_{\mathrm{GC}}$) as a function of time ($t$) over the past $\sim 3.2$ Gyr.}
\label{fig:orbit}
\end{figure}

Figure~\ref{fig:orbit} shows the integrated Galactic orbit of the cluster. The cluster lies in the outer disk ($R_{\rm GC} \approx 13$~kpc) and has a low eccentricity ($e < 0.1$). To properly interpret the influence of this dynamical environment on the BSS population, it is necessary to distinguish between internal cluster dynamics and the external Galactic tidal field \citep{Yontan2023}.

It is generally established that isolated binary evolution, specifically, mass transfer and stellar mergers, is the dominant formation channel for BSSs in OCs \citep{Bisht2026b}. Unlike the dense cores of globular clusters, where dynamical interactions and direct stellar collisions are frequent, the characteristic stellar density of an open cluster like Berkeley~18 is far too low to produce a significant number of collisional BSSs \citep{Leigh2007, Ferraro2012}. Therefore, the size of the BSS population relies strictly on the availability of primordial binary systems that can eventually undergo Roche lobe overflow or merge.

While internal density dictates the physical mechanism of BSS formation, the cluster's Galactic orbit plays a fundamental role in preserving the progenitor binaries. Independent studies on binary fractions demonstrate that star clusters residing in the inner Galactic disk, or those moving on highly eccentric orbits, are subjected to strong and frequent tidal perturbations. These external tidal forces can efficiently strip the cluster of its wider binary systems over time, significantly reducing the overall binary fraction \citep{Marks2011, Parker2014}. Conversely, the low-eccentricity orbit of Berkeley~18 in the outer Galactic disk is consistent with evolution in a relatively benign tidal environment with minimal destructive perturbations. This stable macroscopic environment favors the long-term survival of its binary population, potentially maintaining a reservoir of interacting systems that eventually evolve into the BSSs we observe today.

\section{Photometric Variability in Berkeley 18}
\label{sec:varability}

We investigated the photometric variability of stars within the projected radius of Berkeley 18 using TESS time-series observations, primarily to search for variability signatures among BSS candidates that could constrain binarity or pulsational behaviour. No statistically significant variability is detected among the BSS candidates within the sensitivity limits of the available TESS data, indicating that any variability, if present, is below the detection threshold. This non-detection does not exclude the presence of binary companions, as potential variability may remain undetected due to limited sensitivity, observational noise, or incomplete temporal sampling in the available TESS data. A total of five variable stars were identified in the cluster field. Membership assessment based on $Gaia$ DR3 astrometry indicates that only one of these variables is a high-probability cluster member, while the remaining four objects, including two eclipsing binaries, are classified as non-members. Their detailed properties, including Gaia DR3 source identifiers, stellar parameters, variability periods, and classifications, are summarized in Table~\ref{tab:stellar_param}. The stellar parameters, such as effective temperature, luminosity, and radius, were adopted from the Gaia DR2 and Gaia DR3 catalogues.
We analyzed TESS light curves to investigate the variability characteristics of these stars. The variability periods were determined using the Lomb–Scargle periodogram \citep{lomb1976least, scargle1982studies}, which is particularly well suited for analyzing unevenly sampled time-series data. The resulting periods for each star are listed in Table~\ref{tab:stellar_param}. Figure~\ref{fig: phase_fold} illustrates the analysis, where the top panel shows the original TESS light curve, the middle panel presents the phase-folded light curve, and the bottom panel displays the corresponding Lomb–Scargle periodogram with the False Alarm probability (FAP) level of 0.1$\%$.

\subsection{Cluster Member Variable: SPB Star}

\begin{figure*}
    \centering
    \includegraphics[width=5.5cm,height=7cm]{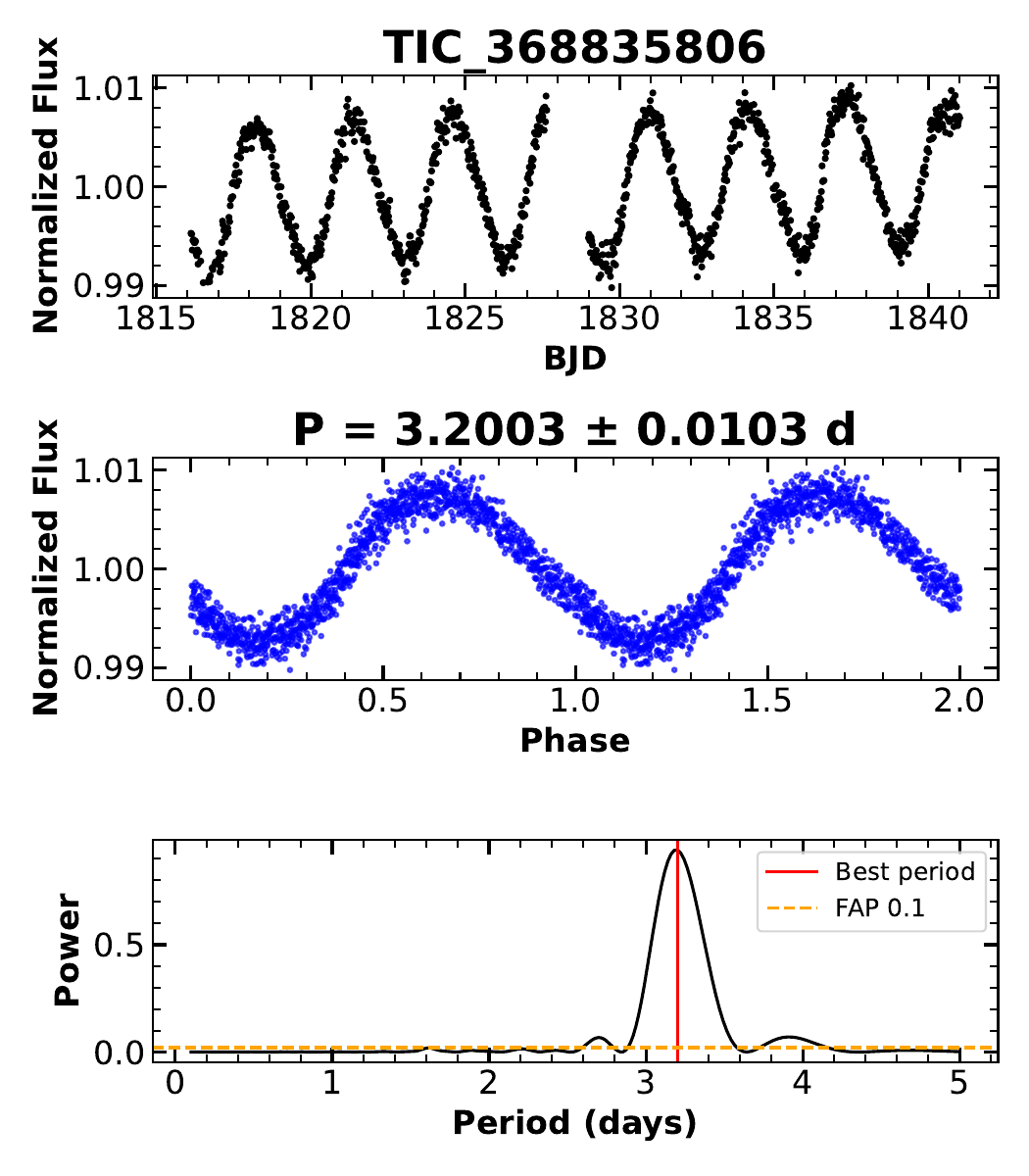}
    \includegraphics[width=5.5cm,height=7cm]{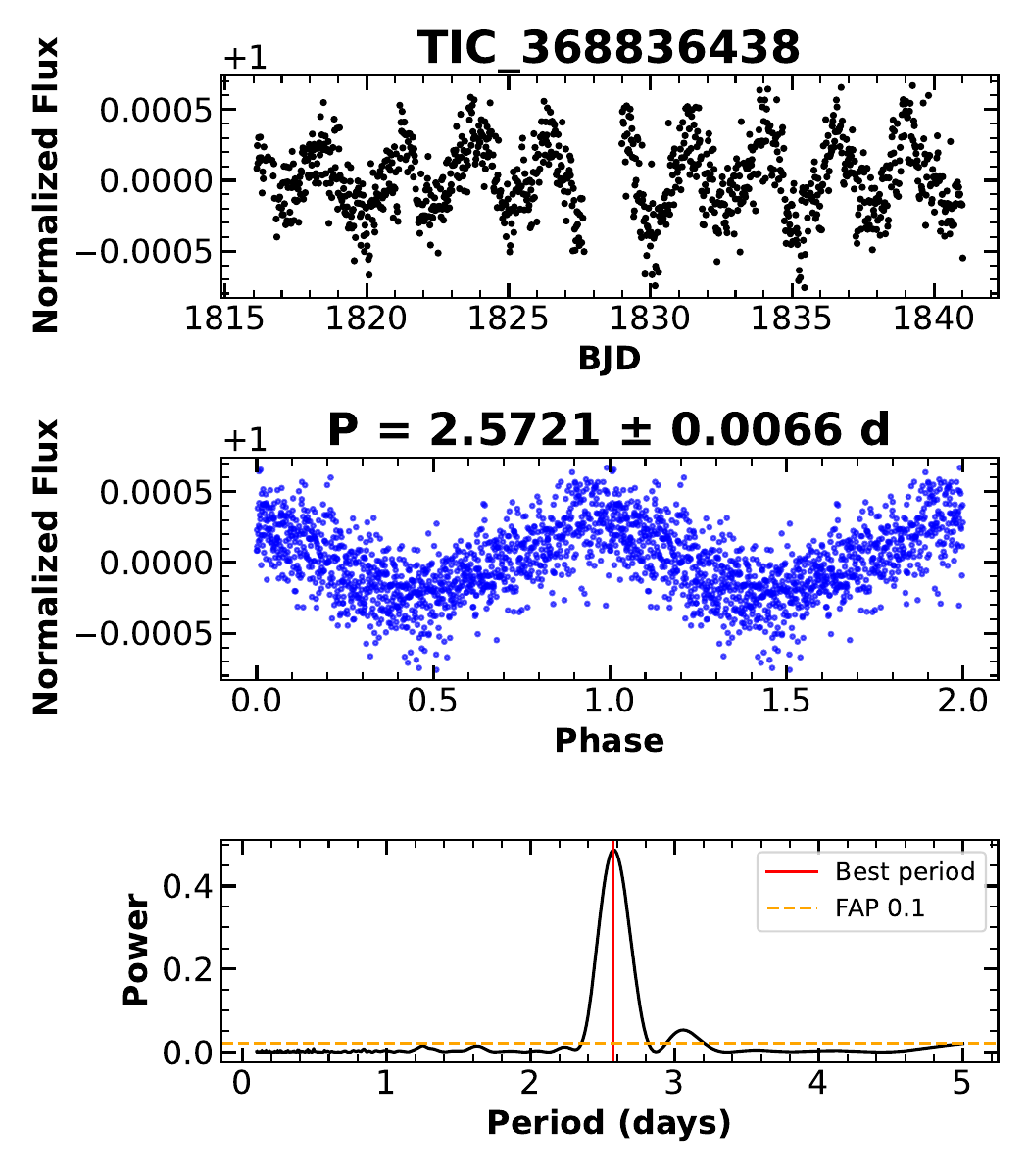}
    \includegraphics[width=5.5cm,height=7cm]{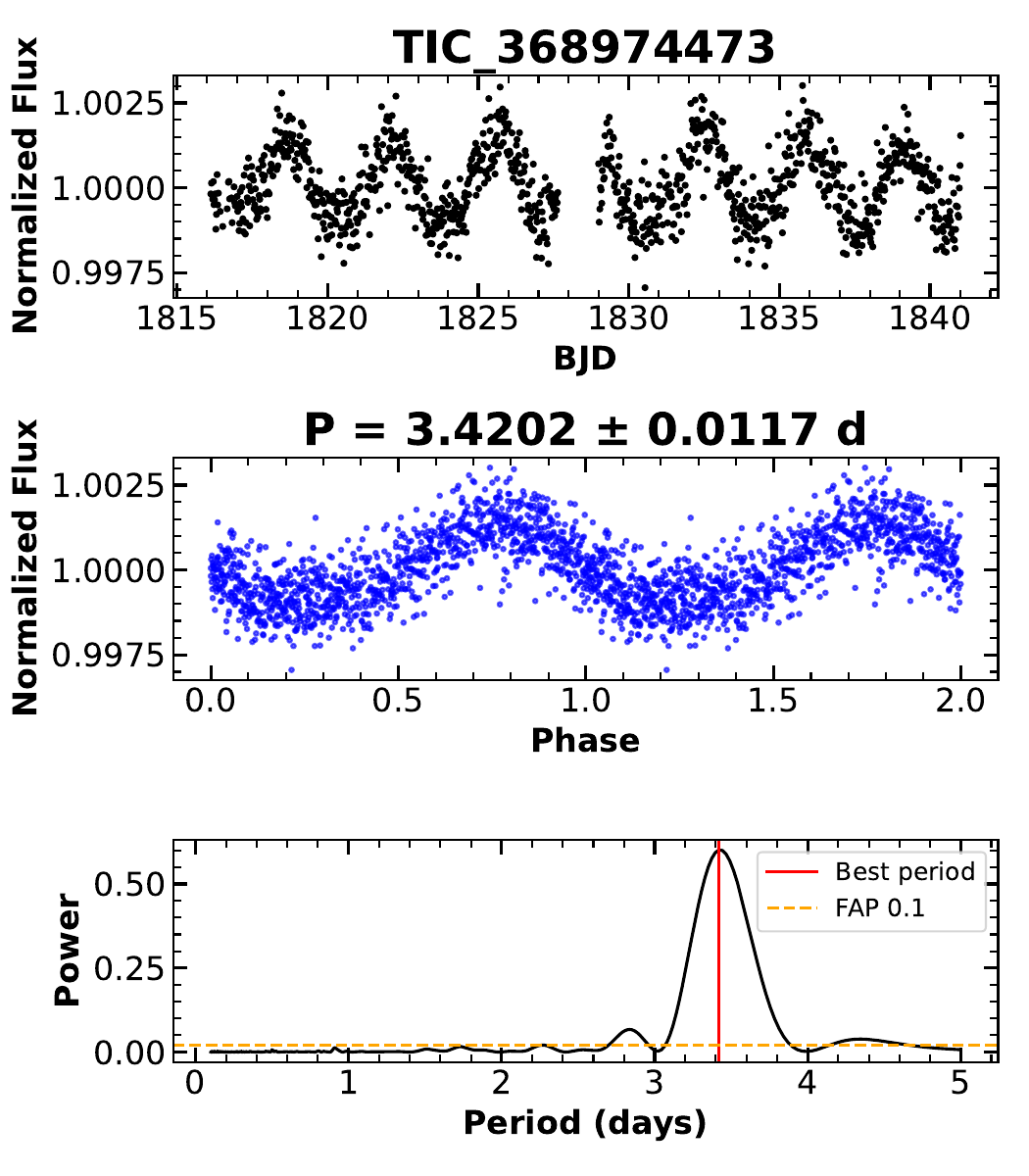}    
    \caption{TESS light curves and periodogram analysis of the identified variable stars in Berkeley 18. For each target (columns), three panels are shown: top—normalized TESS light curve as a function of BJD; middle—phase-folded light curve using the best-fit period, with the sinusoidal model overplotted in blue; bottom—Lomb–Scargle periodogram. The vertical red line marks the best-fit period, while the horizontal dashed line indicates the adopted false alarm probability (FAP = 0.1) significance threshold. The derived periods are labeled in each panel. All three sources exhibit significant periodic signals above the adopted FAP threshold, confirming their variability within the sensitivity limits of the TESS data.}
    \label{fig: phase_fold}
\end{figure*}

\begin{figure}
\centering
\includegraphics[width=0.9\linewidth]{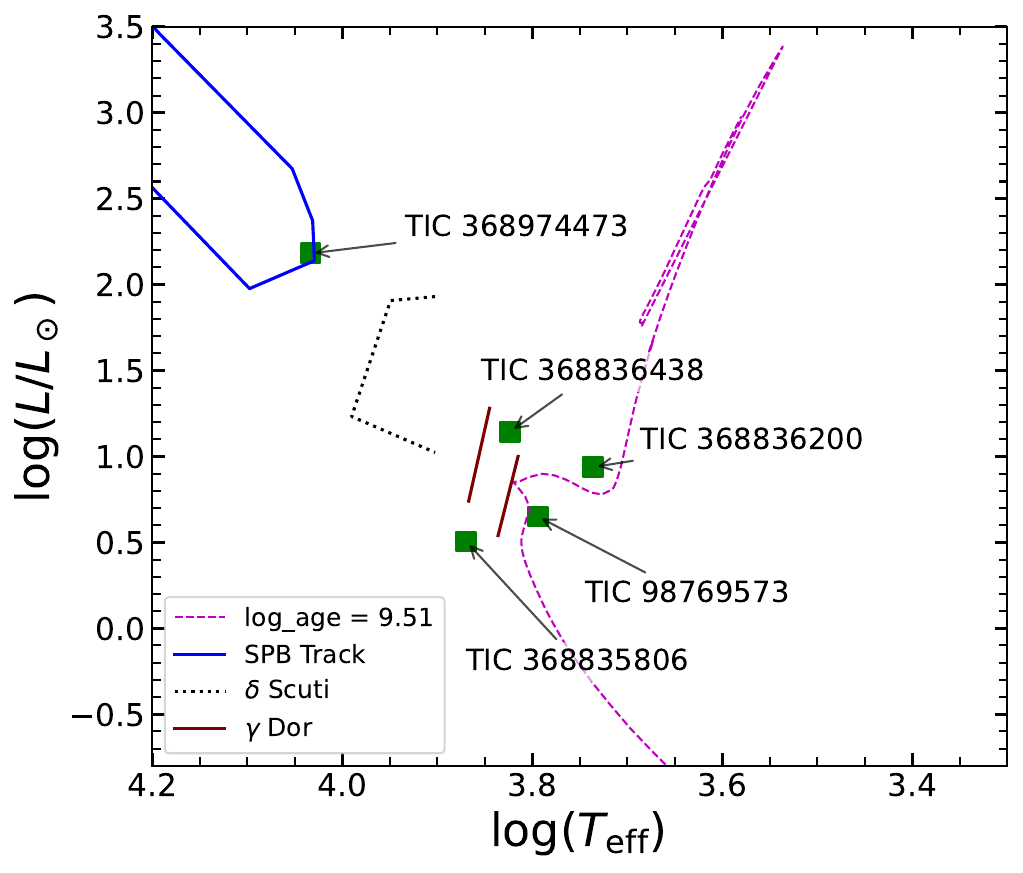} 
\caption{Hertzsprung-Russell diagram showing the locations of the identified variable stars in the cluster field. The filled symbols correspond to the classified variables, with their positions determined from the derived effective temperatures and luminosities. The plotted evolutionary tracks and isochrones correspond to different ages and evolutionary models, including PARSEC isochrones \citep{Bressan2012} and instability-strip boundaries \citep{dupret2005convection}, as indicated in the legend. The instability regions for $\delta$ Scuti, $\gamma$ Dor, and SPB-type variables are also shown for reference. The distribution of the variables across these regions helps constrain their evolutionary stages and potential pulsational nature.}
\label{fig: HR}
\end{figure}

Slowly pulsating B-type (SPB) stars occupy the upper main-sequence region of the H–R diagram and exhibit multi-periodic, non-radial g-mode pulsations driven by the $\kappa$-mechanism \citep{fedurco2020pulsational}. Their typical pulsation periods range from about half a day to several days \citep{stankov2005catalog}. In our analysis, we identify one cluster-member variable, Gaia DR3 208770393097986560 (TIC 368836438; middle panel of Fig.~\ref{fig: phase_fold}), with a derived variability period of $P = 3.42016 \pm 0.12453$ days. The stellar parameters obtained for this source are $\log(T_{\rm eff}) = 4.033 \pm 2.392$, $\log(L/L_{\odot}) = 2.181 \pm 1.239$, and $R = 3.56 \pm 0.178~R_{\odot}$ (Table~\ref{tab:stellar_param}), corresponding to an effective temperature of $\sim 1.08 \times 10^{4}$ K.  The observed period lies well within the expected range for SPB stars and is significantly longer than typical $\delta$ Scuti pulsations ($P < 0.3$ d), supporting a g-mode pulsation origin \citet{waelkens1993slowly}. Furthermore, its position in the H–R diagram places it above the main-sequence turn-off and within the region consistent with the SPB instability strip. The relatively high luminosity and enlarged radius further indicate that the star is more evolved than typical main-sequence stars in the cluster. Taken together, the period, stellar parameters, and HRD location strongly support the classification of this object as an SPB variable. The corresponding instability strip is marked by the blue line in Fig.~\ref{fig: HR}.




\begin{table*}
\centering
\caption{Stellar parameters of selected Gaia DR3 sources. The table lists the Gaia DR3 source identifiers, TESS ID, effective temperatures ($T_{\mathrm{eff}}$), luminosities ($L/L_{\odot}$), stellar radii ($R_{\odot}$), variability periods, membership status, and variability type.}
\begin{tabular}{lccccccc}
\hline
$Gaia$ DR3 Source ID & TESS ID & log($T_{\rm eff}$) & log($L/L_\odot$) & Radius ($R_\odot$) & Period (d) & Membership & Type \\
\hline

208560283300009728 & TIC 98769573  &  3.735 $\pm$ 2.301 & 0.938 $\pm$ 0.285 & 3.31 $\pm$ 0.230 & 8.38078 $\pm$ 0.12845 &  F   & EA \\
208753900425617408 & TIC 368835806 &  3.869 $\pm$ 2.579 & 0.505 $\pm$ 0.480 & 1.09 $\pm$ 0.150 & 3.20033 $\pm$ 0.01028 &  F    &  Misc   \\
208556499427938048 & TIC 368836438 &  3.823 $\pm$ 1.939 & 1.144 $\pm$ 0.615 & 3.19 $\pm$ 0.350 & 2.57211 $\pm$ 0.00664  &  F   &  $\gamma$ Dor  \\
208770393097986560 & TIC 368974473 &  4.033 $\pm$ 2.392 & 2.181 $\pm$ 1.239 & 3.56 $\pm$ 0.178 & 3.42016 $\pm$ 0.12453  &  M  & SPB   \\
208536708218638208 & TIC 68836200  &  3.794 $\pm$ 2.103 & 0.650 $\pm$ 0.230 & 1.99 $\pm$ 0.270 & 1.25354 $\pm$ 0.01174  & F   &  EA \\
 
\hline
\end{tabular}
\label{tab:stellar_param}
\end{table*}

\subsection{Field Variables and Eclipsing Binaries}

Among the four non-member variables, two objects (Gaia DR3 IDs 208560283300009728 and 208536708218638208) are identified as EA-type eclipsing binaries, with periods of $8.38 \pm 0.13$ d and $1.25 \pm 0.01$ d, respectively. We have classified these two stars for the first time as Algol-type binary stars. Their light curves exhibit well-defined primary and secondary minima, consistent with detached binary systems, indicating that they are typical field binaries rather than interacting systems. The remaining two sources include Gaia DR3 ID 208753900425617408, showing low-amplitude variability with a period of $3.20 \pm 0.01$ d, and Gaia DR3 ID 208556499427938048, identified as a $\gamma$ Dor-type pulsator with a period of $2.57 \pm 0.01$ d. Although these objects are not cluster members, they provide a useful comparison sample and demonstrate the reliability of Gaia-based membership selection. The absence of eclipsing binaries among confirmed cluster members further suggests that the observed variability within Berkeley 18 is dominated by intrinsic stellar processes, which is important for interpreting the nature of BSSs.

\subsection{Implications for Blue Straggler Formation}

Although only one variable star is confirmed as a cluster member, its properties provide meaningful insight into the formation of blue straggler stars in Berkeley 18. Its location in the blue straggler region, combined with its high luminosity, enlarged radius, and SPB-type variability, strongly suggests a rejuvenated nature, likely the result of past binary interaction. The presence of such variability is consistent with expectations from mass-transfer or merger scenarios, in which binary evolution alters the star's internal structure and evolutionary state. These results therefore support binary evolution as a plausible formation channel for at least a fraction of the blue straggler population in this cluster.

\subsection{Binary Modeling with PHOEBE}

To further investigate the nature of the identified eclipsing systems, we performed detailed light-curve modeling using the PHOEBE (Physics of Eclipsing Binaries) code. The modeling was carried out for two representative systems, TIC 98769573 and TIC 368836200, using the phase-folded TESS light curves. The best-fitting solutions were obtained by simultaneously optimizing key orbital and stellar parameters, including the mass ratio ($q$), orbital inclination ($i$), effective temperatures ($T_{\mathrm{eff,1}}$, $T_{\mathrm{eff,2}}$), surface potentials ($\Omega_1$, $\Omega_2$), and stellar radii.

To investigate the physical and geometrical properties of the eclipsing binary TIC 98769573 and TIC 368836200, we modelled its TESS light curve using the PHOEBE 1.0 software package \citep{prvsa2005computational}. The orbital period was fixed to the value derived from our period analysis (1.25354, 8.38078 d). Determining the mass ratio ($q = m_2/m_1$) is a crucial initial step in photometric light-curve modelling. Since no radial velocity measurements are available for this system, we adopted the commonly used q-search technique to estimate the mass ratio (e.g., \citep{li2023five, panchal2025exploring, Kuldeep2025, belwal2026time}. In this method, light-curve solutions are generated over a range of trial mass ratios, while the remaining system parameters are optimized simultaneously. The most reliable solution is the one that minimizes $\chi^2$. For the q-search analysis, the updated ephemeris was used to transform the light curve from BJD to phase–flux space. The effective temperature of the primary component was kept fixed during the fitting process. As contact binaries are expected to have convective outer envelopes, the gravity-darkening coefficients were fixed at $g_1 = g_2 = 0.32$, and the bolometric albedos were set to $A_1 = A_2 = 0.5$. The surface potentials of both components were assumed to be equal ($\Omega_1 = \Omega_2$), consistent with a contact configuration. The synchronicity parameters were fixed at unity, and a circular orbit was assumed. The limb-darkening coefficients were automatically interpolated within PHOEBE using the square-root law based on the tabulations of \citet{van1993new}.

The PHOEBE modelling indicates that both TIC 98769573 and TIC 368836200 are detached eclipsing binary systems. The short orbital period of TIC 98769573 (1.25 days) suggests a compact configuration with possible tidal interaction, while TIC 368836200, with a period of 8.38 days, represents a wider system. The derived mass ratios indicate that TIC 98769573 is a strongly unequal mass binary, whereas TIC 368836200 shows more comparable component masses. The effective temperatures suggest F–G type stellar components in both systems. The relatively large radii of the components, particularly in TIC 368836200, indicate possible evolutionary expansion beyond the zero-age main sequence. The Roche potentials confirm that both systems remain detached, indicating no significant ongoing mass transfer. Such detached eclipsing binaries are important laboratories for testing stellar evolution models because they allow direct determination of fundamental stellar parameters such as masses, radii, and luminosities \citep{prvsa2005computational, lee2010claur, torres2010accurate}.



In comparison, TIC 368836200 shows a higher mass ratio ($q \approx 0.66$) and a longer orbital period ($P \approx 8.38$ days), indicating a more balanced binary configuration. The inclination ($i \sim 78^\circ$) also supports the presence of deep eclipses. The relatively similar surface potentials and effective temperatures of the two components further confirm the system's contact nature.

Taken together, the derived stellar parameters, including masses, radii, and luminosities, are consistent with evolved binary systems undergoing mass exchange. In particular, the combination of short orbital periods, contact configurations, and mass ratios supports the interpretation that these systems have experienced significant angular momentum loss and mass transfer.

The identified eclipsing systems are primarily classified as detached (EA-type) binaries, consistent with the PHOEBE modelling results discussed above. As most of these variables are non-members and only one system is associated with the cluster, the present variability analysis does not directly constrain BSS formation in Berkeley~18. Nevertheless, the detection and characterization of these systems highlight the effectiveness of time-domain observations in identifying binary systems along the line of sight.

\begin{figure}
    \centering
    \includegraphics[width=1\linewidth]{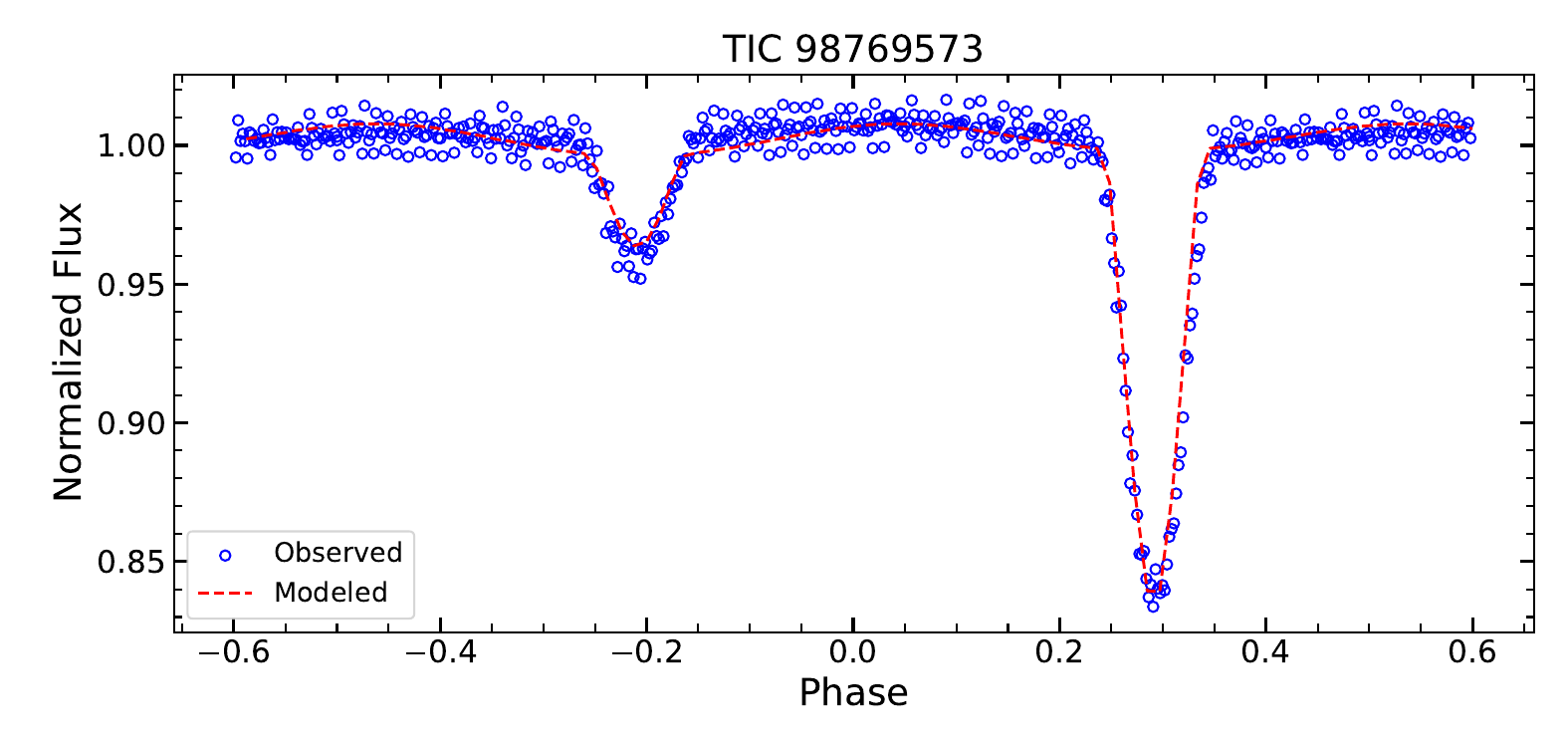}\\
    \includegraphics[width=1\linewidth]{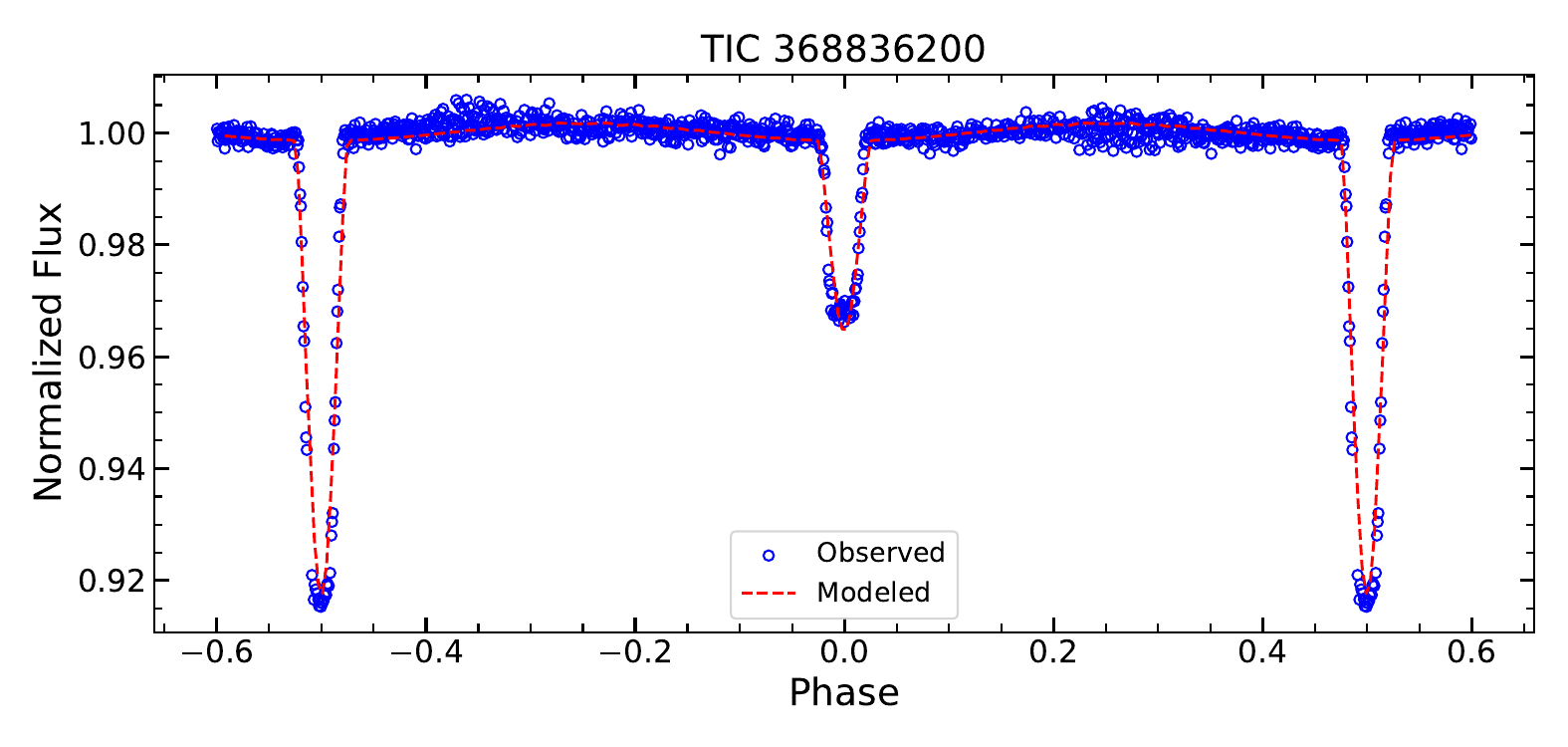}
    \vspace{-0.5cm}
    \caption{Phase–folded TESS light curves of the two eclipsing binary systems TIC 98769573 and TIC 368836200 detected in the direction of Berkeley 18. Blue circles represent the observed normalized flux, and red dashed curves show the best–fit models obtained using PHOEBE.}
    \label{fig:phoebe_model}
\end{figure}

\begin{table}
    \centering
    \caption{Orbital and physical parameters of TIC 98769573 and TIC 368836200 derived from binary modeling. Subscripts 1 and 2 denote the primary and secondary components, respectively. Both systems are modeled as detected binaries ($\Omega_1 = \Omega_2$).}
    \label{tab:phoebe_table}
    \hspace{-1cm}
    \begin{tabular}{c|c|c}
    \hline
    \hline
    Parameter     &   TIC 98769573  &  TIC 368836200  \\
    \hline
     q   &  0.290   &     0.66   \\
     i($^{o}$)   &  71.75   &   78.44     \\
     period (d)   & 1.25354    &  8.38078      \\
     $\Omega_{1}$   & 3.6778    &  7.4300      \\
     $\Omega_{2}$   & 3.6778    &  7.4300       \\
     $T^{\rm eff}_{1}$   & 6773    &  6370      \\
     $T^{\rm eff}_{2}$   & 6090    &  5693      \\ 
     L$_{1}$(L$_{\odot}$)   &  2.400   &   3.1721     \\
     L$_{2}$(L$_{\odot}$)   & 2.262    &   2.7905     \\     
     M$_{1}$(M$_{\odot}$)   & 1.6551    &  0.5463      \\
     M$_{2}$(M$_{\odot}$)   & 0.4799    &  0.3605      \\  
     R$_{1}$(R$_{\odot}$)   & 1.877    &   2.4862     \\
     R$_{2}$(R$_{\odot}$)   & 0.7889    &  1.7731      \\  

    \hline     
    \end{tabular}

\end{table}

Overall, no statistically significant photometric variability within the sensitivity and time baseline of the TESS data. The typical photometric precision of the TESS full-frame images is $\sim$1--3 mmag, with a temporal baseline of $\sim$27 days per sector. This implies that variability with amplitudes below $\sim$0.01 mag or with periods longer than the TESS observing window may remain undetected. The absence of detectable variability, therefore, primarily constrains short-period, high-amplitude variability but does not rule out binarity. Many post-mass-transfer systems are expected to exhibit low-amplitude, long-period, or inclination-dependent variability that may fall below the detection threshold or be poorly sampled. Additionally, once active mass transfer ceases, such systems may evolve into relatively photometrically stable configurations. Therefore, the lack of observed variability is consistent with a population of evolved BSSs formed through binary interactions, where current photometric signatures are weak or absent. Further constraints on the binary nature of these systems will require high-precision, long-baseline photometric monitoring and dedicated spectroscopic observations.

\section{Conclusions and Summary}
\label{sec:conclusion}

We present a detailed multiwavelength analysis of the old open cluster Berkeley 18, combining optical and infrared photometry with Gaia~DR3 astrometry to constrain its structural, evolutionary, and dynamical properties, as well as to characterise its blue straggler star (BSS) population. The main findings of this study are summarised as follows:

\begin{enumerate}
\item Using a GMM applied to Gaia~DR3 astrometric data, we identified 798 probable members in Berkeley 18. The mean proper-motion components derived for the cluster are in excellent agreement with those reported by \citet{Cantat-Gaudin20} and \citet{Hunt2023}, confirming the reliability of the adopted membership selection.

\item Isochrone fitting to the Gaia color--magnitude diagram yields an age of $3.2 \pm 0.2$~Gyr and a heliocentric distance of $5.01^{+0.75}_{-0.55}$~kpc. These results place Berkeley~18 among the older open clusters located in the outer Galactic disk, suggesting that it has undergone dynamical evolution over several Gyr.

\item The structural parameters derived from King-profile fitting give a core radius of $r_c = 6.91^{+0.91}_{-0.73}$ arcmin and a tidal radius of $r_t = 13.23^{+0.44}_{-0.43}$ arcmin. These values suggest a moderately concentrated yet dynamically evolved system, consistent with expectations for old open clusters that have survived long-term mass loss in the Galactic environment.

\item We identify 24 BSS candidates above the main-sequence turn-off. Their spectral energy distributions, modeled using multiwavelength photometry, reveal a wide range of physical properties with effective temperatures of $T_{\rm eff} \sim 6000$--$8500$~K, radii of $R \sim 1.4$--$5.7\,R_\odot$, and luminosities spanning $L \sim 3$--$38\,L_\odot$. This diversity indicates that the BSS population is not uniform but comprises multiple evolutionary stages, ranging from relatively unevolved objects to more inflated, evolved systems.

\item The spatial distribution of BSSs shows a mild central concentration compared to RGB stars, although the overall distributions partially overlap. The low value of the $A^{+}_{\mathrm{rh}}$ parameter further indicates a weak degree of mass segregation. In addition, the extremely low stellar collision-rate proxy strongly disfavors a predominantly collisional origin for the BSS population. Together, these results provide a consistent picture in which dynamical interactions are inefficient in Berkeley~18, and strongly support binary evolution as the dominant formation mechanism of BSSs in this low-density environment.

\item Orbital integrations using \texttt{galpy} within the \texttt{MWPotential2014} Galactic model reveal that Berkeley 18 follows a nearly circular thin-disk orbit with low eccentricity ($e \approx 0.08$) and modest vertical excursion ($Z_{\max} < 1$~kpc). Such a dynamically stable Galactic environment is consistent with a relatively low level of disruptive tidal interactions and may favor the long-term survival of primordial binary systems.

\item A search for circumstellar material was conducted using broadband mid-infrared photometry. Although automated SED fitting initially suggested apparent long-wavelength (WISE W3/W4) excesses in two candidates, BSS~11 and BSS~17, a detailed visual inspection indicates that these signals are likely artifacts due to background contamination and source blending. Independent SPHEREx spectrophotometry further confirms the absence of any near- or mid-infrared excess up to $5~\mu$m, supporting the conclusion that the WISE long-wavelength excesses are not physical. Consequently, we find no robust evidence for circumstellar dust around these candidates, highlighting the importance of careful visual validation when identifying disk candidates in crowded cluster environments.

\item We analyzed TESS time-series data and identified five variable stars in the cluster field, including eclipsing binaries and periodic variables. Gaia~DR3 membership indicates that only one is a probable cluster member, while the others are field contaminants. Light-curve modeling using PHOEBE was performed for selected systems. Notably, none of the 24 BSS candidates show statistically significant variability at the TESS sensitivity limits, consistent with the absence of strong ongoing binary-interaction signatures.

\end{enumerate}

This study presents a comprehensive multiwavelength investigation of BSSs in the old open cluster Berkeley~18, providing robust observational constraints on their formation and evolutionary pathways. The combination of a low $A^{+}$ parameter, an extremely low stellar collision-rate proxy, the absence of reliable mid-infrared excess, and the non-detection of significant photometric variability collectively indicates that dynamical interactions are inefficient in this system. These results strongly support a scenario in which the BSS population is predominantly formed through binary evolution, most likely as evolved post-mass-transfer systems rather than actively interacting binaries. 
Berkeley~18 thus serves as an important benchmark for studying the interplay between stellar dynamics, binary evolution, and long-term cluster survival in the outer Galactic disk. Future high-resolution spectroscopic observations and radial-velocity monitoring will be essential for directly confirming binarity and further constraining the evolutionary histories of these systems.

\section*{Acknowledgements}

We sincerely thank Dr. Andrea Veronica Ahumada for the careful evaluation of our manuscript and for the valuable comments and suggestions that helped us improve the quality and clarity of the paper. Ing-Guey Jiang acknowledges support from the National Science and Technology Council (NSTC), Taiwan, under grants NSTC 113-2112-M-007-030 and NSTC 114-2112-M-007-029. This work has used data from the European Space Agency (ESA) mission \textit{Gaia}, processed by the \textit{Gaia} Data Processing and Analysis Consortium (DPAC). Funding for the DPAC has been provided by national institutions participating in the \textit{Gaia} Multilateral Agreement. This work has also used data from the SPHEREx project, under a contract from the NASA Goddard Space Flight Center to the California Institute of Technology.

This research has used the VOSA tool, developed under the Spanish Virtual Observatory project, funded by MCIN/AEI/10.13039/501100011033/ through grant PID2020-112949GB-I00. VOSA has been partially updated with funding from the European Union's Horizon 2020 Research and Innovation Program under grant agreement No. 776403 (EXOPLANETS-A).

\section*{Data Availability}

The data underlying this article are available in public repositories. 
\textit{Gaia} DR3 astrometric and photometric data, together with multiwavelength photometry from 2MASS, WISE, Pan-STARRS1, and SkyMapper, were obtained from the VizieR catalogue access tool and the \textit{Gaia} archive. 
TESS full-frame image data are available from the Mikulski Archive for Space Telescopes (MAST) at \href{http://dx.doi.org/10.17909/t9-nmc8-f686}{10.17909/t9-nmc8-f686}. 
SPHEREx spectrophotometric data were retrieved using the publicly available SPHEREx \footnote{\url{https://irsa.ipac.caltech.edu/applications/spherex/tool-spectrophotometry}} Spectrophotometry Tool based on Level 2 calibrated data products. 
Finder chart images were obtained from the STScI Digitized Sky Survey (DSS). The derived data products and analysis codes supporting this study are available from the corresponding author upon reasonable request.

\bibliographystyle{mnras}
\bibliography{References}

\appendix
\section{SED Fits of Blue Straggler Stars}
\label{Appendix_SED}

\begin{figure*}
\centering
\includegraphics[width=0.99\linewidth]{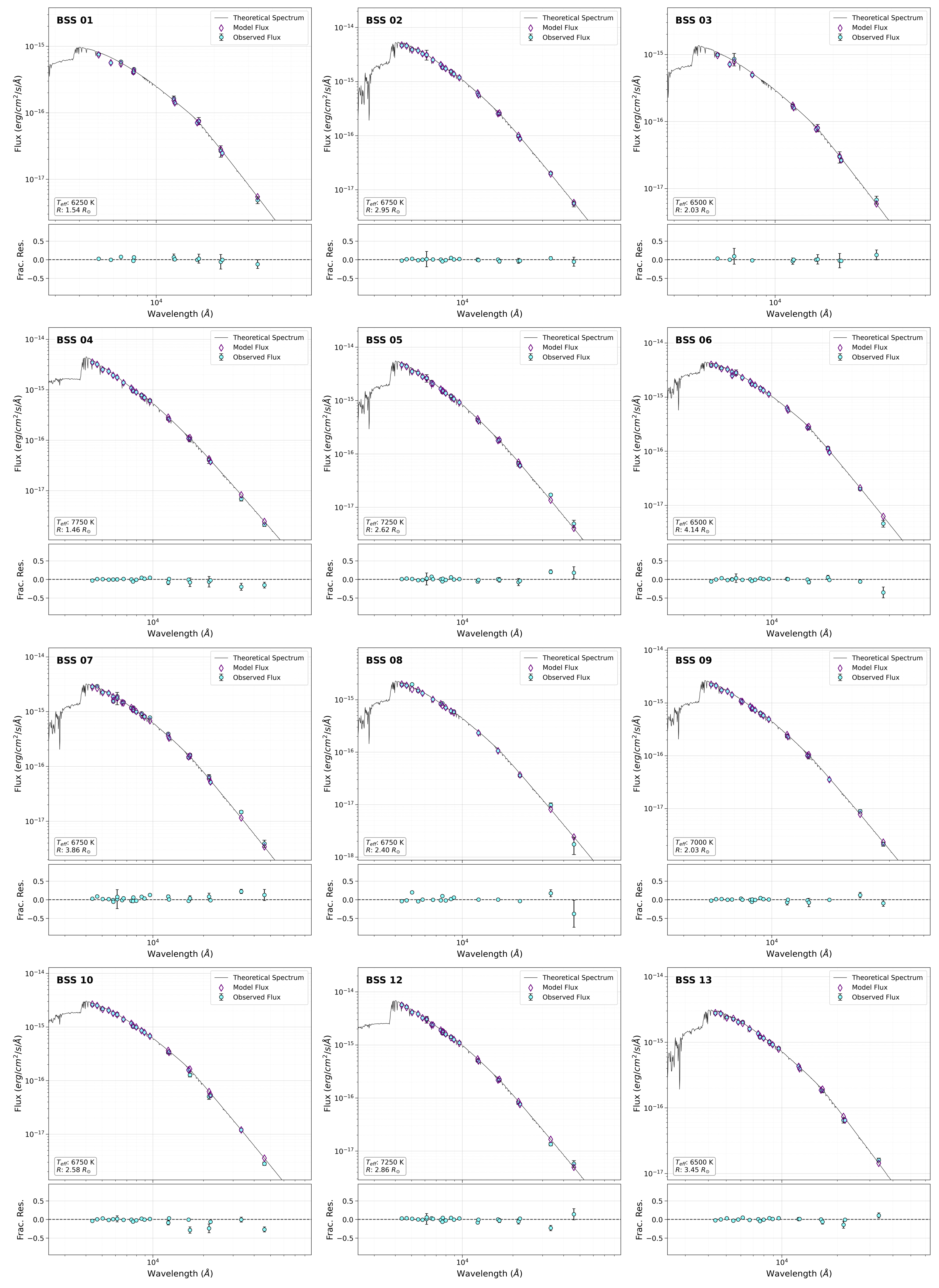}
\caption{Spectral energy distributions of blue straggler stars.}
\label{fig:appendix_sed_1}
\end{figure*}

\begin{figure*}
\centering
\includegraphics[width=0.99\linewidth]{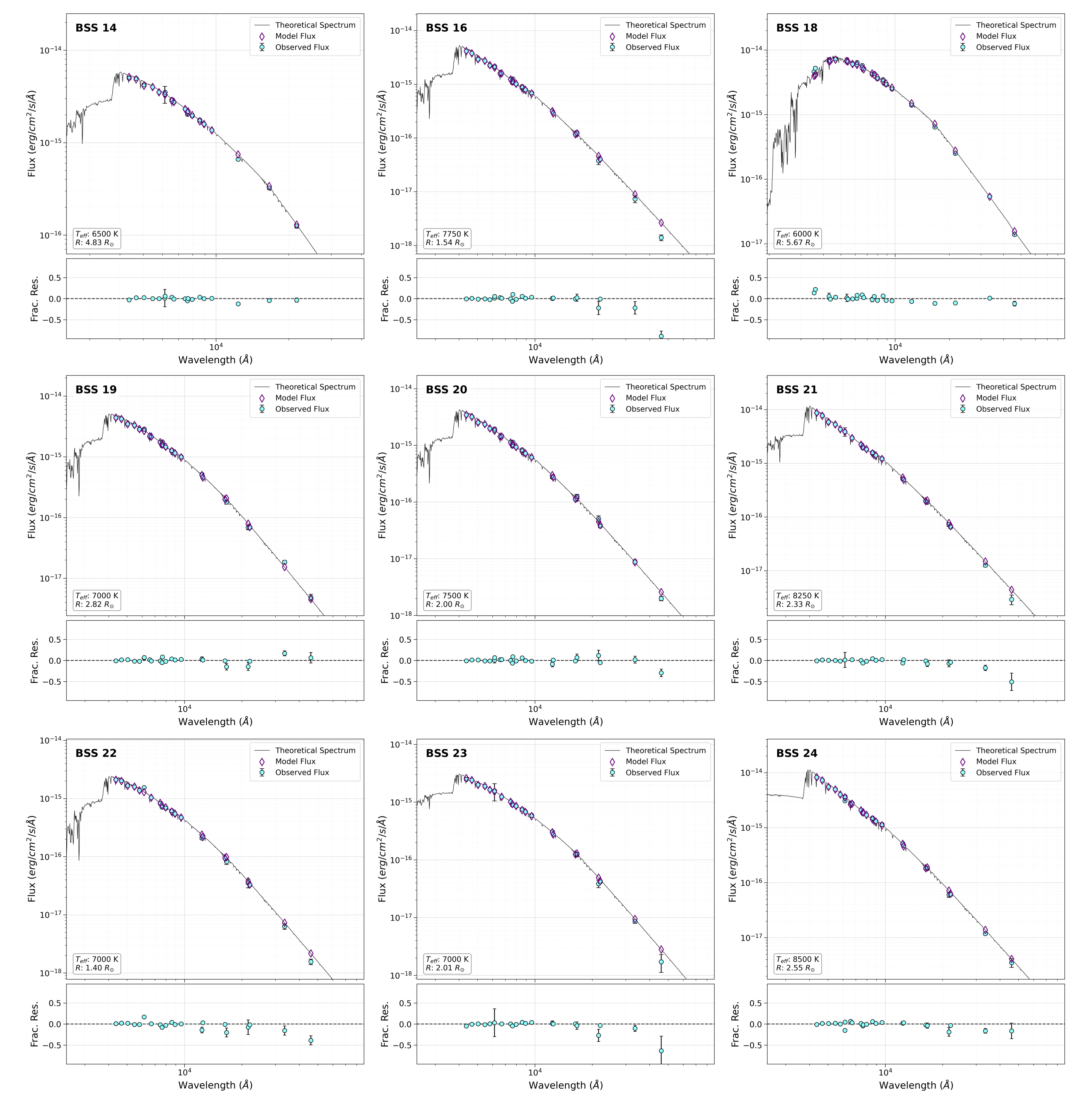}
\caption{(Continued.)}
\label{fig:appendix_sed_2}
\end{figure*}

\begin{figure*}
\centering
\includegraphics[width=0.48\linewidth]{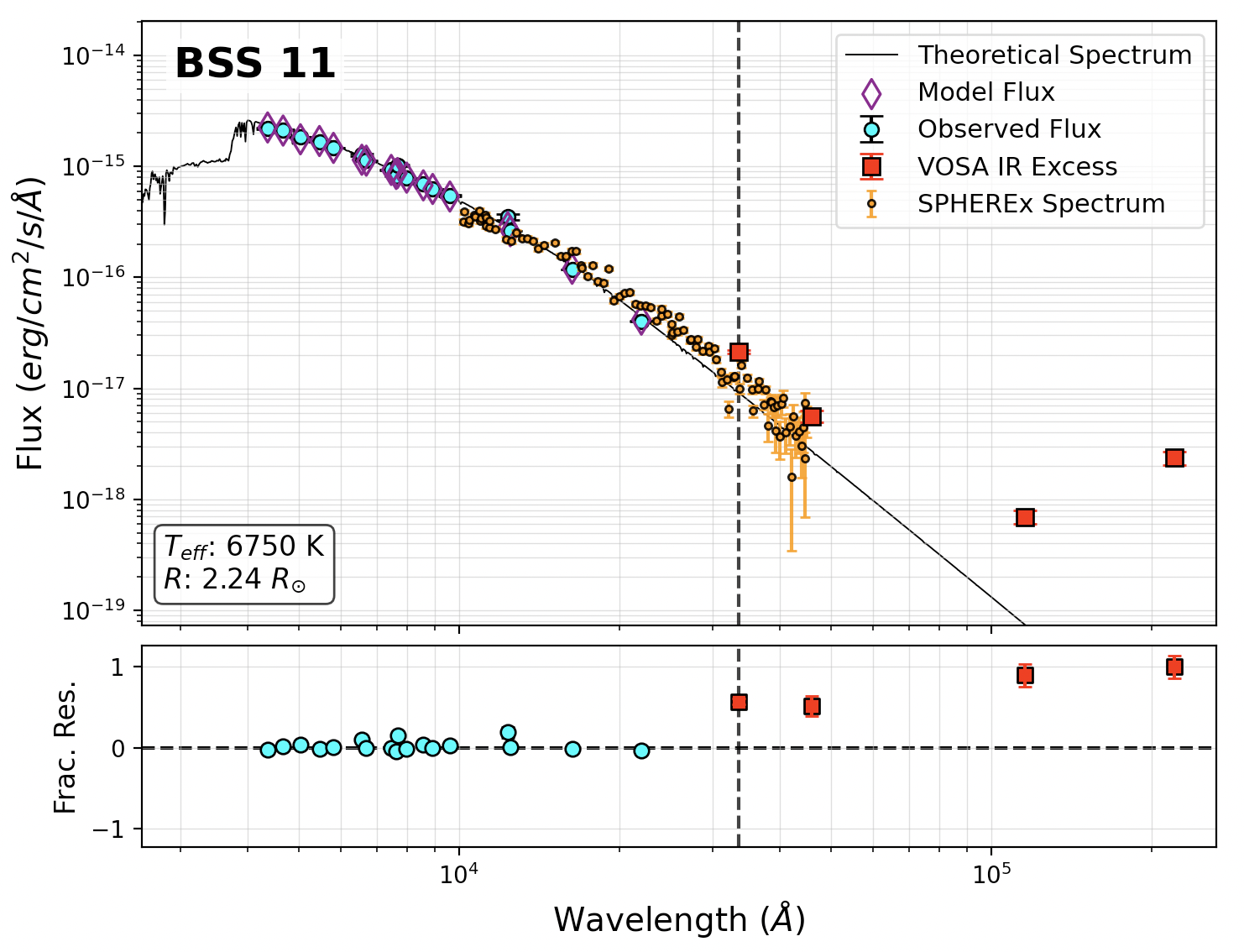}
\includegraphics[width=0.48\linewidth]{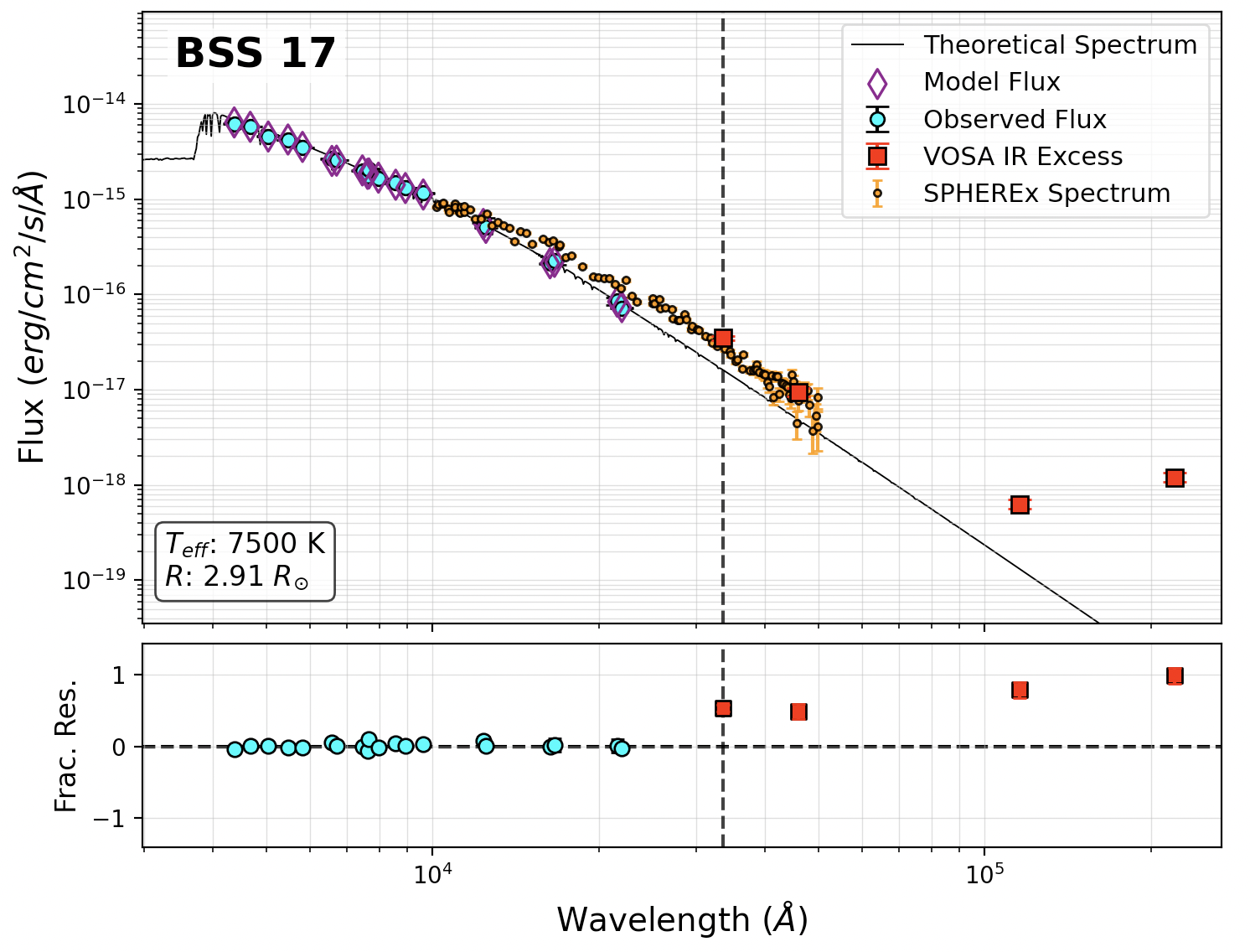}
\caption{SEDs for BSS~11 (left) and BSS~17 (right) obtained with VOSA, overlaid with SPHEREx spectrophotometric data. Black solid lines show the best-fitting stellar-atmosphere models, with purple diamonds indicating the model fluxes integrated over the filter curves. Observed VOSA photometry is shown as cyan circles, while red circles mark the WISE mid-IR measurements. SPHEREx spectrophotometry is represented by orange circles. The vertical dashed line marks the onset of the VOSA IR excess. The bottom panels display the fractional residuals of the observed fluxes relative to the stellar model.}
\label{fig:IR_Excess}
\end{figure*}

\bsp	
\label{lastpage}
\end{document}